\documentclass[
  journal=pasa,
  manuscript=research-paper,
  year=2026,
  volume=YY,
]{cup-journal}

\usepackage{amsmath}
\usepackage{amssymb,microtype,siunitx,booktabs}
\usepackage{graphicx}
\usepackage{subcaption}
\usepackage{url}
\usepackage{multirow}
\usepackage{rotating}
\usepackage{xcolor}
\usepackage{adjustbox}
\usepackage{tabularx}
\title{Constraining FRB Microstructure with Polarised Shot Noise}

\author{J.~C.~F.~Balzan}
\affiliation{International Centre for Radio Astronomy Research (ICRAR), Curtin University, Bentley, WA 6102, Australia}
\email[J.~C.~F.~Balzan]{joel.balzan@icrar.org}

\author{A.~Bera}
\affiliation{ASTRON, Netherlands Institute for Radio Astronomy, Postbus 2, 7990 AA Dwingeloo, Netherlands}
\alsoaffiliation{International Centre for Radio Astronomy Research (ICRAR), Curtin University, Bentley, WA 6102, Australia}

\author{C.~W.~James}
\affiliation{International Centre for Radio Astronomy Research (ICRAR), Curtin University, Bentley, WA 6102, Australia}

\author{B.~W.~Meyers}
\affiliation{Australian SKA Regional Centre (AusSRC), Curtin University, Kent Street, Bentley, WA 6102, Australia}
\alsoaffiliation{International Centre for Radio Astronomy Research (ICRAR), Curtin University, Bentley, WA 6012, Australia}

\received{dd Mmm YYYY}
\revised{dd Mmm YYYY}
\accepted{dd Mmm YYYY}
\published{22 September 202X}


\begin{document}

\begin{abstract}
We present \texttt{FIRES}, a polarised shot-noise (PSN) framework that models fast radio burst (FRB) dynamic spectra as the incoherent superposition of Gaussian microshots with varying polarisation angles (PAs). Applied to the CRAFT bursts FRB~20191001A and FRB~20240318A, \texttt{FIRES} can reproduce key spectro-polarimetric behaviours seen in these data: scattering suppresses PA variability on the trailing edge, while the leading edge preferentially retains intrinsic structure when sufficient signal-to-noise is present. We quantify this behaviour using the PA variance ratio $\mathcal{R}_\psi$ and explore the joint plane of measured linear polarisation fraction $\Pi_L$ versus PA variance to identify allowed regions of microshot number $N$, intrinsic PA dispersion $\sigma_\psi$, and intrinsic linear fraction $\Pi_{L,0}$ at fixed signal-to-noise. This restricts these combinations permitted within the adopted shot-noise framework. For FRB~20191001A, the data are consistent with an extended parameter space, reflecting degeneracies between intrinsic PA structure, microshot superposition, scattering, finite sampling, noise, and the assumed microshot-property distributions. FRB~20240318A occupies a more restricted region, favouring fewer microshots and larger intrinsic PA dispersion. By combining an emission-mechanism-independent forward-modeling framework with minimal assumptions and observational constraints, \texttt{FIRES} facilitates qualitative and quantitative exploration of how microshot superposition, scattering, finite sampling, and noise can shape observed FRB polarimetry under the PSN model.

\end{abstract}

\section{INTRODUCTION}
Fast Radio Bursts (FRBs) are transient radio signals characterised by their short duration and high brightness temperatures~\citep{Lorimer2007}. The FRB population is diverse, exhibiting various properties such as apparent ``one-off'' events or strong repeaters \citep[see][for recent reviews]{2019A&ARv..27....4P,2022A&ARv..30....2P}, highly variable polarisation angles (PAs), linear and circular polarisation fractions, rotation measures (RMs), as well as complex temporal and spectral profiles with multiple peaks and varying widths \citep[e.g.,][]{2024ApJ...964..131S,2025ApJ...979..160S,2025arXiv250517497S}. The study of these structures is essential for deciphering FRB generation mechanisms and the media through which they propagate.

FRBs show diverse PA behaviour. Only one FRB is well fit by a rotating vector model~\citep[RVM;][]{2025Natur.637...43M}, a characteristic of a rotating neutron star, with circumstantial evidence for two more~\citep[][]{2024ApJ...969L..29B}. Attempts to fit RVMs to a large population of repeater bursts shows that some are well fit, however, their results are inconsistent with each other~\citep[][]{2025ApJ...988..175L}. Other FRBs show a variety of PA swings~\citep[][]{2020Natur.586..693L}, or apparently flat PA profiles~\citep[e.g.,][]{2024ApJ...968...50P}. The Canadian Hydrogen Mapping Experiment (CHIME) and the Deep Synoptic Array 110 (DSA-110) previously catagorised FRBs by their spectro-temporal morphology~\citep{2021ApJ...923....1P,2025ApJ...979..160S}, the smooth PA evolution measured across the burst envelope~\citep{2024ApJ...968...50P} and their polarisation fractions~\citep{2024ApJ...964..131S}. These are highly dependent on S/N, time resolution, and propagation effects and leads to arbitrary classifications as suggested by~\cite{2025arXiv250517497S}. Studying burst-envelope-scale PA behaviour may be useful for identifying rotational signatures or probing large-scale, spatially averaged magnetic field structure, and can inform us of potential progenitors; however, it may provide limited insight into the rapid local plasma processes that generate the emission, which may arise either in the magnetosphere~\citep[e.g.,][]{2018MNRAS.477.2470L} or in more distant outflow/shock regions~\citep[e.g.,][]{2014MNRAS.442L...9L}.

\cite{2025arXiv250517497S} present high-time-resolution (HTR) polarisation properties of 35 FRBs from the Commensal Real-time ASKAP Fast Transients (CRAFT) Incoherent-Sum Survey~\citep{2025PASA...42...36S}. They move away from previous classifications, showing that all FRBs likely have multiple components and have intrinsically variable PAs. However, measuring these temporal structures can be influenced by several factors, including the intrinsic properties of the source and its environment, and particularly scattering along the line of sight. When sampled with HTR, FRBs with little-to-no scattering and high signal-to-noise (S/N) can exhibit complex, narrow intrinsic widths~\citep[nanoseconds--microseconds;][]{2022NatAs...6..393N} and highly variable PAs within the burst envelope \citep[e.g.,][]{2022MNRAS.514.5866G,2023NatAs...7.1486S,2025arXiv250517497S}. Observations of these structures can place stringent constraints on the size of the emitting region and emission mechanisms of FRBs, as well as the various propagation media \citep[e.g.][]{2020MNRAS.498..651B}.

The idea of radio pulses being comprised of multiple smaller ``shots'' was first suggested by~\cite{1975ApJ...197..185R} in the amplitude-modulated noise model for pulsars. This idea was extended by~\cite{1976ApJ...210..780C} who implemented polarisation variations to explain short-time scale polarisation structures, the so-called ``polarised shot noise'' (PSN) model. Many further developments have since been made to the model to account for stochastic processes~\citep[e.g.,][]{1998ApJ...505..921M,2006ApJ...645..551M,2011MNRAS.418.1258O}. The most famous application of the shot-models is to the Crab pulsar, which displays nanosecond duration shots (nanoshots) with PA microstructure~\citep[][]{2007ApJ...670..693H,2016ApJ...833...47H}. \cite{2022NatAs...6..393N} and \cite{2025arXiv250517497S} similarly note that their FRB samples have structures resembling Crab nanoshots. 

Of the Crab pulsar's many components, single pulses in the main pulse and interpulse exhibit similar polarimetric nanostructure to FRBs. They often appear substantially depolarised at time resolutions of order tens of $\mu$s to ms, but at nanosecond time resolution and with minimal scattering they resolve into 100\% polarised nanoshots with random PAs~\citep{2015ApJ...802..130H,2016ApJ...833...47H}. Though, the PA variability of these very bright pulses observed at HTR can be affected by self-noise and small-number statistics when the averaging is insufficient \citep{2009ApJ...694.1413V}. \cite{2016ApJ...833...47H} show that unresolved microbursts generally have highly disordered PA profiles, consistent with shots having random PAs. Occasionally their PAs show ordered swings (see their Figure~4) which may trace underlying plasma or magnetic field structures from the emission environment. They also show that microbursts are in fact unresolved collections of narrowband nanoshots, with some demonstrating substructures of timescales less than 4\,ns (see their Figure~8). This behaviour is consistent with expectations from the PSN model. 
Giant pulses (GPs) are a subclass of single pulses with luminosities several orders of magnitude greater than the average pulses, a distinct energy distribution (power law), and a narrow emission phase window (compared to the average pulse profile). In terms of duration and energetics, Crab GPs and their underlying nanoshot mechanism are the most analogous phenomena to FRBs~\citep[e.g.,][]{2022NatAs...6..393N}. These Crab phenomena each show intense PA variability on nanosecond to microsecond timescales~\citep[e.g.,][]{2010A&A...524A..60J,2016ApJ...833...47H} and a variety of spectro-temporal morphologies, on durations similar to FRBs observed at very-high time resolution. 

A natural question is whether depolarisation in FRBs may arise from the superposition of many microshots, analogous to the behaviour of Crab microbursts. Other contributors include depolarisation from scattering, Faraday rotation through turbulent media, or the intrinsic emission mechanism itself. Disentangling these possibilities requires controlled forward modelling of FRB microstructure and polarisation at HTR. However, the current sample of high-S/N, low-scatter FRBs with nanosecond-microsecond resolution remains limited, motivating the need for realistic simulations.

Several existing packages offer pulsar-signal simulation capabilities \citep[e.g.,][]{2004PASA...21..302H,2021JOSS....6.2757H}, but these are generally unsuited to FRBs because they model emission profiles with simple Gaussian components and lack polarisation-analysis tools. In this paper, we introduce the Fast, Intense Radio Emission Simulator (\texttt{FIRES}),\footnote{\url{https://github.com/JoelBalzan/FIRES}} a Python package for simulating and analysing FRB dynamic spectra under the PSN model. We describe \texttt{FIRES}, and demonstrate how \texttt{FIRES} can reproduce key observed FRB properties,including PA variance, depolarisation, and linear polarisation fractions, show how microshot superposition can naturally generate the diversity of polarimetric structures seen in FRBs, and show to what extent FRB parameters can be constrained.
\section{MODELLING DYNAMIC SPECTRA}
\label{sec:model}

\begin{table*}[ht!]
\centering
  \caption{Model parameters used in \texttt{FIRES} as assumed inputs for representative simulations.}
\label{tab:params}
\begin{tabular}{llll}
\toprule
Parameter & Description & FRB 20191001A & FRB 20240318A \\
\midrule
$N$ & Number of microshots & $100$ & $100$ \\
$A_i$ (Jy) & Peak amplitude of microshots;\ power law, $\alpha=-3$ & $[0.15,15]$ & $[0.015,1.5]$ \\
$t_i$ (ms) & TOA of $i$th microshot, $t_i\sim\mathcal{N}(t_0,\sigma_t)$ 
  & $\mathcal{N}(t_0,\sigma_t)$ & $\mathcal{N}(t_0,\sigma_t)$ \\
$w_i$ (ms) & Microshot FWHM, sampled via Eq.~(\ref{eq:width}) & sampled & sampled \\
$W_0$ (ms) & Envelope FWHM & $0.50$ & $0.06$ \\
$w_{\min}$ (fraction of $W_0$) & Min.\ fractional microshot width & $0.05$ & $0.05$ \\
$w_{\max}$ (fraction of $W_0$) & Max.\ fractional microshot width & $0.20$ & $0.20$ \\
$t_0$ (ms) & Envelope reference TOA & fixed & fixed \\
$\psi_0$ (deg) & Mean intrinsic PA & $20$ & $0$ \\
$\sigma_\psi$ (deg) & microshot PA scatter & $30$ & $20$ \\
$\Pi_{L,0}$ & Intrinsic frac.\ linear pol. & $0.99$ & $0.98$ \\
$\Pi_{V,0}$ & Intrinsic frac.\ circular pol. & $-0.05$ & $-0.17$ \\
$\nu_\tau$ (MHz) & Reference freq.\ for scattering \& scintillation & $1000$ & $1000$ \\
$\tau_{0}$ (ms) & Scattering timescale at $\nu_\tau$ 
  & $1.78^a \pm 0.04$ & $0.128^a \pm 0.005$ \\
$\alpha$ & Scattering index 
  & $-4.85^a \pm 0.3$ & $-3.22^a \pm 0.005$ \\
$\nu_s$ (MHz) & Scintillation decorrelation bandwidth & $2.05 \pm 0.06$ & $4^a \pm 0.2$ \\
$t_s$ (s) & Scintillation timescale & $300$ & $300$ \\
$N_\mathrm{im}$ & Number of images in scattering disk & $5000$ & $5000$ \\
$\theta_{\mathrm{lim}}$ (dimensionless)$^{b}$ & Angular limit of scattering disk & $3$ & $3$ \\
SEFD (Jy) & System equivalent flux density & $1.2$ & $1.4$ \\
$\delta \nu$ (MHz) & Channel bandwidth & $1$ & $1$ \\
$\delta t$ (ms) & Time resolution & $0.023$ & $0.003$ \\
\bottomrule
\end{tabular}
\medskip\\
\raggedright $^a$ Data from \citet{2025arXiv250517497S}.\\
\raggedright $^b$ Defined in units of the characteristic scattering angle (see Section~\ref{subsec: scint}).\\
\end{table*}

Following the idea that all FRBs intrinsically have multiple components~\citep{2025arXiv250517497S}, we model the dynamic spectra of FRBs as a superposition of Gaussian microshots with different PAs~\citep[][]{1976ApJ...210..780C} using \texttt{FIRES}. Each microshot is defined by its amplitude, time of arrival (TOA), width, PA, and polarisation fractions, with the latter specifying that each microshot is partially elliptically polarised. For our initial model, we assume that each microshot has a Gaussian intensity profile with a peak amplitude sampled from a power law distribution, consistent with Crab GP energetics~\citep[e.g.,][]{2019MNRAS.490L..12B}. In principle, a degeneracy exists between the number of shots and their amplitudes: an alternative model could instead assume a constant number of shots whose amplitudes vary to reproduce the envelope. The amplitude distribution of Crab nanoshots does seem consistent with a power law~\citep[e.g., Figure 8 from][]{2016ApJ...833...47H}, however, this has not yet been quantitatively constrained (see~\ref{subsec: amp dist} for a comparison of amplitude distribution power laws).

We further assume that their TOAs are distributed normally, and we ignore frequency dependence in intrinsic power. Given the high linear polarisation of FRBs and commonly assumed neutron star magnetospheric emission models~\citep[e.g,][]{2018MNRAS.477.2470L}, we adopt a picture in which the emitting magnetic field is predominantly ordered with only small perturbations. We therefore sample microshot PAs from a normal distribution. Each microshot has its full-width half-maximum (FWHM) sampled from a uniform distribution, and all other parameters are fixed. The microshots are then scattered by convolution with a thin-screen scattering response, diffractive interstellar scintillation (DISS) is applied as a multiplicative gain field, and noise is added to create realistic FRB dynamic spectra. \texttt{FIRES} can apply Faraday rotation and PA trends, however, we do not use them here. The list of parameters used in the model are presented in Table~\ref{tab:params}.

The Stokes I dynamic spectrum, comprised of $N$ microshots in time $t$ (ms) and frequency $\nu$ (MHz), is
\begin{align*}
    D_I[t,\nu] = \left[ \sum_{i=1}^{N} A_i \exp\!\left(-\frac{(t - t_i)^2}{2\sigma_{w,i}^2}\right) \right] * h_\nu(t),
    \\
    \sigma_{w,i} = \frac{w_i}{2\sqrt{2\log{2}}}, 
\end{align*}
where $*$ denotes convolution in time with the frequency-dependent, thin-screen scattering response $h_\nu(t)$. Each microshot has a peak amplitude $A_i$~Jy, drawn from a power law distribution with index $\alpha = -3$, and a Gaussian FWHM $w_i$, where $W_0$ (ms) is the envelope FWHM and the fractional microshot width $w_i/W_0$ is drawn from
\begin{equation}\label{eq:width}
    \frac{w_i}{W_0} \sim \mathcal{U}\!\left(w_{\min},\, w_{\max}\right),
\end{equation}
with $w_{\min}$ and $w_{\max}$ the minimum and maximum fractional microshot widths, respectively. $t_i$ is the arrival time of the $i$-th microshot, which is distributed as $t_i \sim \mathcal{N}\!\left(t_0,\, \sigma_t\right)$, where $t_0$ is the arrival time of the envelope, and 
\begin{equation}\label{eq:sig t}
    \sigma_t = \frac{W_0}{2\sqrt{ 2\log{2}}}.
\end{equation}

With per–microshot polarisation angles $\psi_i \sim \mathcal{N}(\psi_0,\sigma_\psi)$, where $\psi_0$ is the mean intrinsic PA and $\sigma_\psi$ is the standard deviation of the intrinsic microshot-to-microshot PA scatter, and fixed intrinsic fractional linear and circular polarisations $\Pi_{L,0}$ and $\Pi_{V,0}$, we write
\begin{align*}
    D_Q[t, \nu] &= D_I[t, \nu] \cdot \Pi_{L,0} \cdot \cos(2\psi_i), \\
    D_U[t, \nu] &= D_I[t, \nu] \cdot \Pi_{L,0} \cdot \sin(2\psi_i), \\
    D_V[t, \nu] &= D_I[t, \nu] \cdot \Pi_{V,0}.
\end{align*}

\subsection{Thin-Screen Scattering}
We apply scattering via a discrete, causal impulse response per frequency channel. The continuous thin–screen scattering response is
\begin{equation*}
    h_\nu(t) \;=\; \frac{1}{\tau_\nu}\,e^{-t/\tau_\nu}\,H(t),
\end{equation*}
with Heaviside step $H(t)$ and timescale $\tau_\nu$. Let $\delta t$ be the time resolution and let
\begin{equation*}
    n \equiv \left\lceil 5\,\frac{\tau_\nu}{\delta t} \right\rceil + 1
\end{equation*}
be the number of kernel samples. Then the implemented kernel is the truncated, sum-normalised exponential
\begin{align*}
    \mathrm{IRF}_{\nu}[k] =& \frac{e^{-k\delta t / \tau_\nu}}
        {\displaystyle \sum_{k=0}^{n-1} e^{-k\delta t / \tau_\nu}} \\
    =& \frac{\bigl(1 - e^{-\delta t / \tau_\nu}\bigr)\, e^{-k\delta t / \tau_\nu}}
           {1 - e^{-n\delta t / \tau_\nu}},
    \qquad k=0,\dots,n-1,
\end{align*}
so that $\sum_{k=0}^{n-1} \mathrm{IRF}_{\nu}[k]=1$ (preserving flux), where $k$ indexes the $k$-th discrete time sample. We zero‑pad each time series on the right by $(n-1)$ samples, perform a linear (FFT) convolution, and truncate to the original timespan; any power beyond the simulated window is discarded.

The frequency-scaled scattering timescale is
\begin{equation}
    \tau_\nu = \tau_{0}\left(\frac{\nu}{\nu_\tau}\right)^{\alpha},
\end{equation}
with reference $\nu_\tau = 1\,\mathrm{GHz}$, frequency dependence $\alpha$, and $\tau_{0}$ the timescale at the reference frequency in milliseconds.

\subsection{Scintillation}\label{subsec: scint}
We include DISS as a multiplicative gain field $G(t,\nu)$ applied to all Stokes vectors. We generate a complex electric field $E(t,\nu)$ using \texttt{ScintillationMaker}\footnote{\url{https://github.com/SprengerT/ScintillationMaker/tree/main}} with four control parameters: the decorrelation bandwidth $\nu_s$, characteristic timescale $t_s$, number of images drawn from the scattering disk $N_\mathrm{im}$, and the (dimensionless) angular extent of the scattering disk $\theta_\mathrm{lim}$, defined in units of the characteristic scattering angle (so that $\theta_{\rm lim} = 3$ corresponds to sampling the disk out to $\sim3\sigma$). In our simulations we set $t_s = 300~\mathrm{s}$, $N_\mathrm{im} = 5000$, and $\theta_\mathrm{lim} = 3.0$ to adequately sample the scattering disk and produce a realistic scintillation pattern while keeping computations manageable. $\nu_s$ is taken from \cite{2025arXiv250517497S} when available, otherwise it is measured by fitting a single Lorentzian in the frequency power spectrum (see \ref{appendix:scint}). The observed intensity gain is
\begin{equation*}
    G(t,\nu) \equiv \frac{\bigl|E(t,\nu)\bigr|^2}{\left\langle \bigl|E(t,\nu)\bigr|^2 \right\rangle_{t,\nu}},
\end{equation*}
so that $\langle G\rangle=1$ over the simulated $(t,\nu)$ grid (preserving mean flux). The scintillated Stokes parameters are then
\begin{equation*}
    D_X^{(\mathrm{DISS})}(t,\nu) = G(t,\nu)\, D_X(t,\nu), \qquad X \in \{I,Q,U,V\}.
\end{equation*}

For the large-batch simulations presented in the main text and appendices, scintillation is omitted to reduce computational cost. As discussed in \ref{appendix:scint}, tests including representative scintillation modulation patterns do not affect the results obtained in this work

\subsection{Noise}
To mimic telescope noise, we add Gaussian noise to the dynamic spectra with
\begin{equation*}
    \varepsilon_X(t,\nu) \sim \mathcal{N}\!\left(0,\, \sigma_X\right),
\end{equation*}
where $\sigma_X$ is the noise level calculated from the radiometer equation,
\begin{equation}\label{eq:rad}
    \sigma_X = \frac{\mathrm{SEFD}}{\sqrt{n_\mathrm{pol}\delta\nu \delta t}}.
\end{equation}
SEFD is the system equivalent flux density in Jy, $n_\mathrm{pol}=2$ is the number of polarisations ``recorded,'' $\delta\nu$ is the frequency resolution in Hz, and $\delta t$ is the time resolution in seconds. For the ASKAP data considered here, this Gaussian approximation is appropriate. At substantially higher time and frequency resolution, however, the noise can become non-Gaussian (e.g. closer to a $\chi^2$ distribution in minimally averaged baseband data), so alternative noise models may be required in future applications of \texttt{FIRES}. The observed Stokes are
\begin{equation*}
    D_X \leftarrow D_X^{(\mathrm{DISS})} + \varepsilon_X, \qquad X \in \{I,Q,U,V\}.
\end{equation*}

\subsection{Polarisation Angles}
We calculate the final PAs and their errors as performed by~\cite{2020MNRAS.497.3335D} and similarly debias the linear polarisation following (the typographically corrected) Equation 11 in \citet[][]{2001ApJ...553..341E}, 
\begin{equation*}
    L_\mathrm{debias} = 
    \begin{cases}
        \sigma_{L}\sqrt{\frac{L_{\text{meas}}}{\sigma_{L}}^{2}-1}, & \dfrac{L_{\text{meas}}}{\sigma_{L}} \ge 1.57 \\[6pt]
        0, & \text{otherwise},
    \end{cases}
\end{equation*}
except using $\sigma_L$ instead of $\sigma_I$ to accommodate cases where $\sigma_I \neq \sigma_Q \neq \sigma_U$. Further, we discard $\psi$ where $L_\mathrm{debias}<2\sigma_{L}$, along with those outside the on-pulse region, the minimum boxcar width that contains 95\% of the flux in the pulse profile. If PA values are in a sequence of length $<5$, then we also discard them.

\section{REPRODUCING REAL FRB STRUCTURES}\label{sec:reproduce real FRB structures}
\begin{figure*}[ht!]
    \centering
    \begin{subfigure}[b]{0.45\textwidth}
        \includegraphics[width=\textwidth]{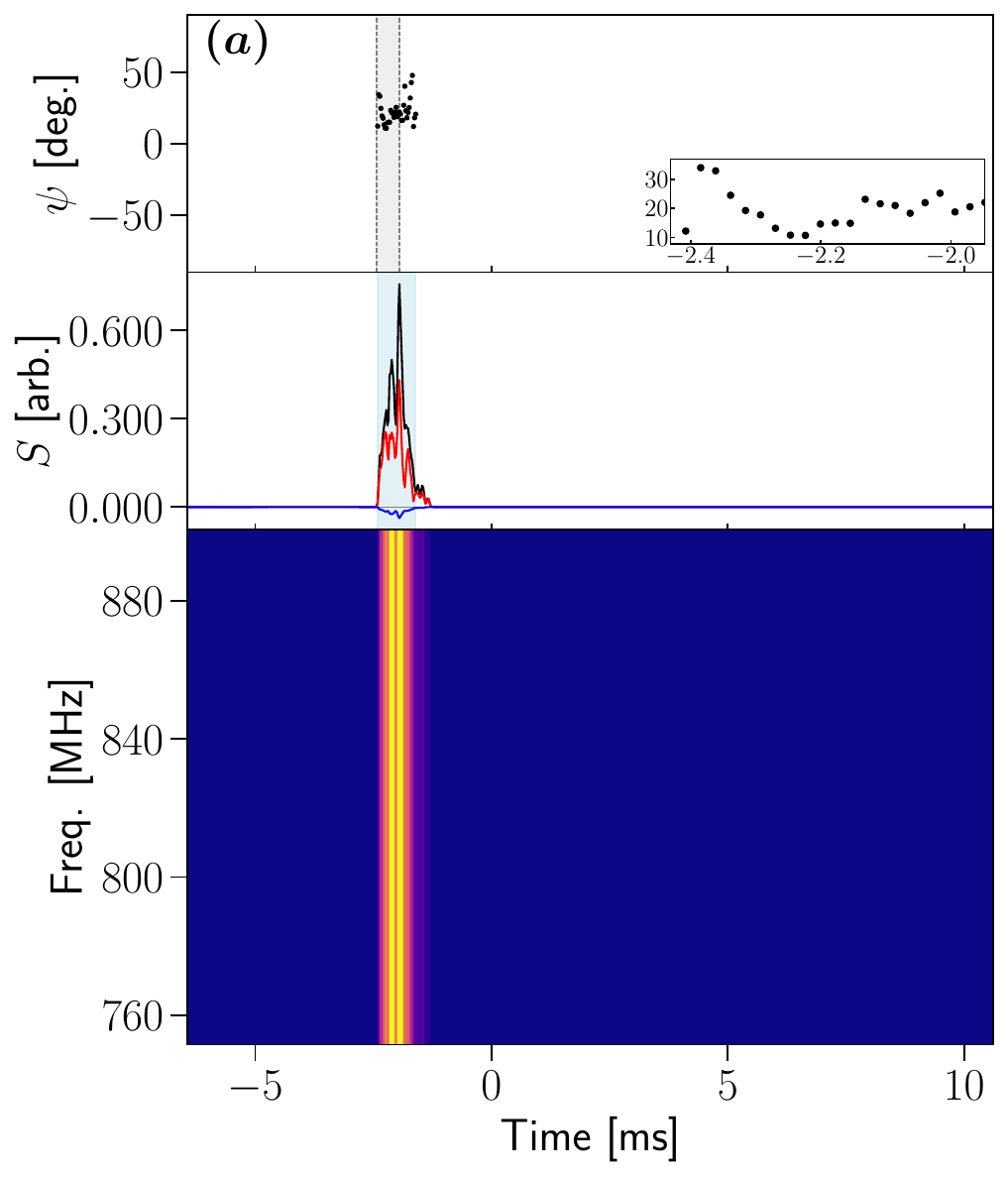}
        \captionlistentry{}\label{fig:191001_ns_nn}
    \end{subfigure}
    \begin{subfigure}[b]{0.45\textwidth}
        \includegraphics[width=\textwidth]{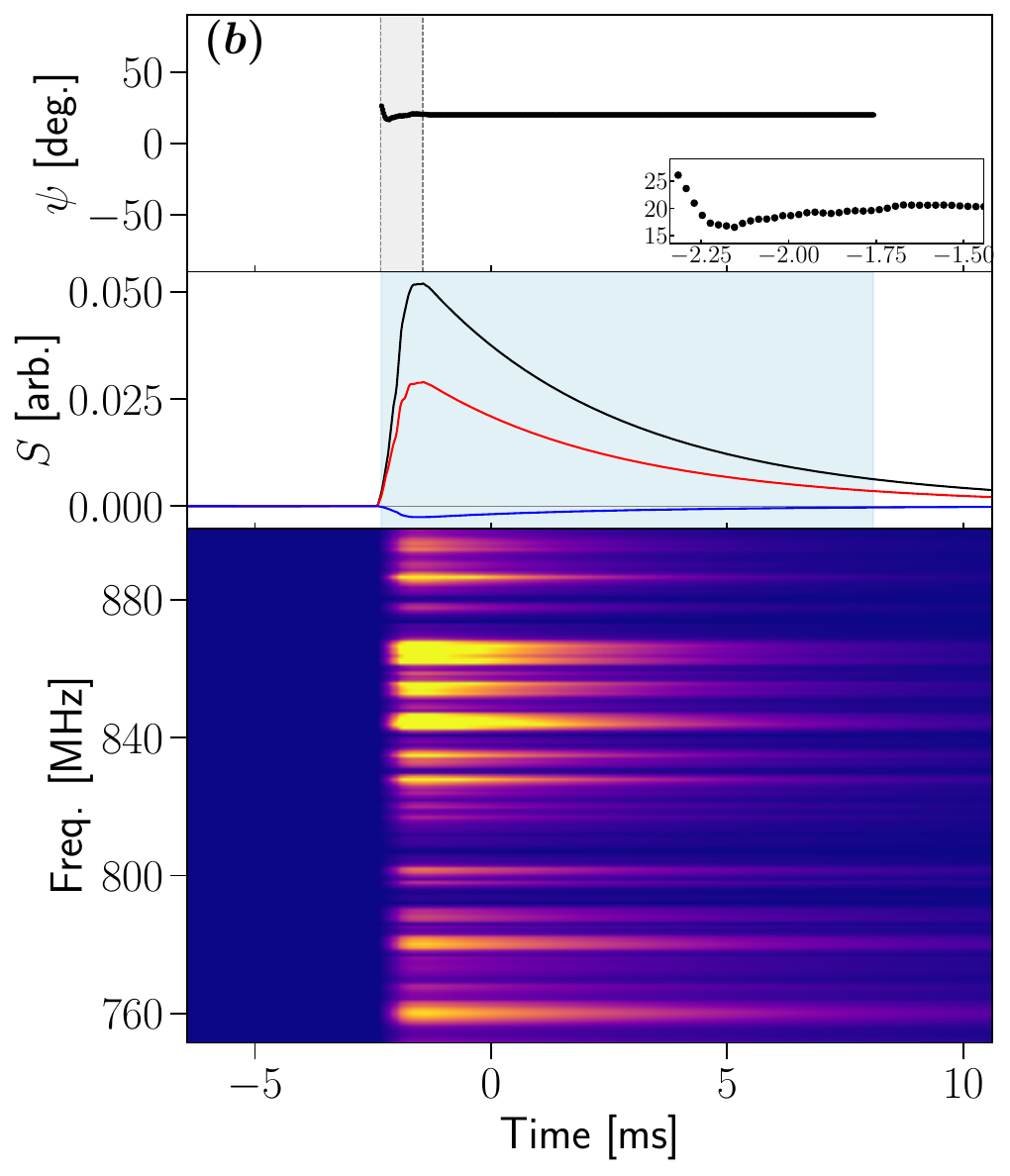}
        \captionlistentry{}\label{fig:191001_ns}
    \end{subfigure}
    \begin{subfigure}[b]{0.45\textwidth}
        \includegraphics[width=\textwidth]{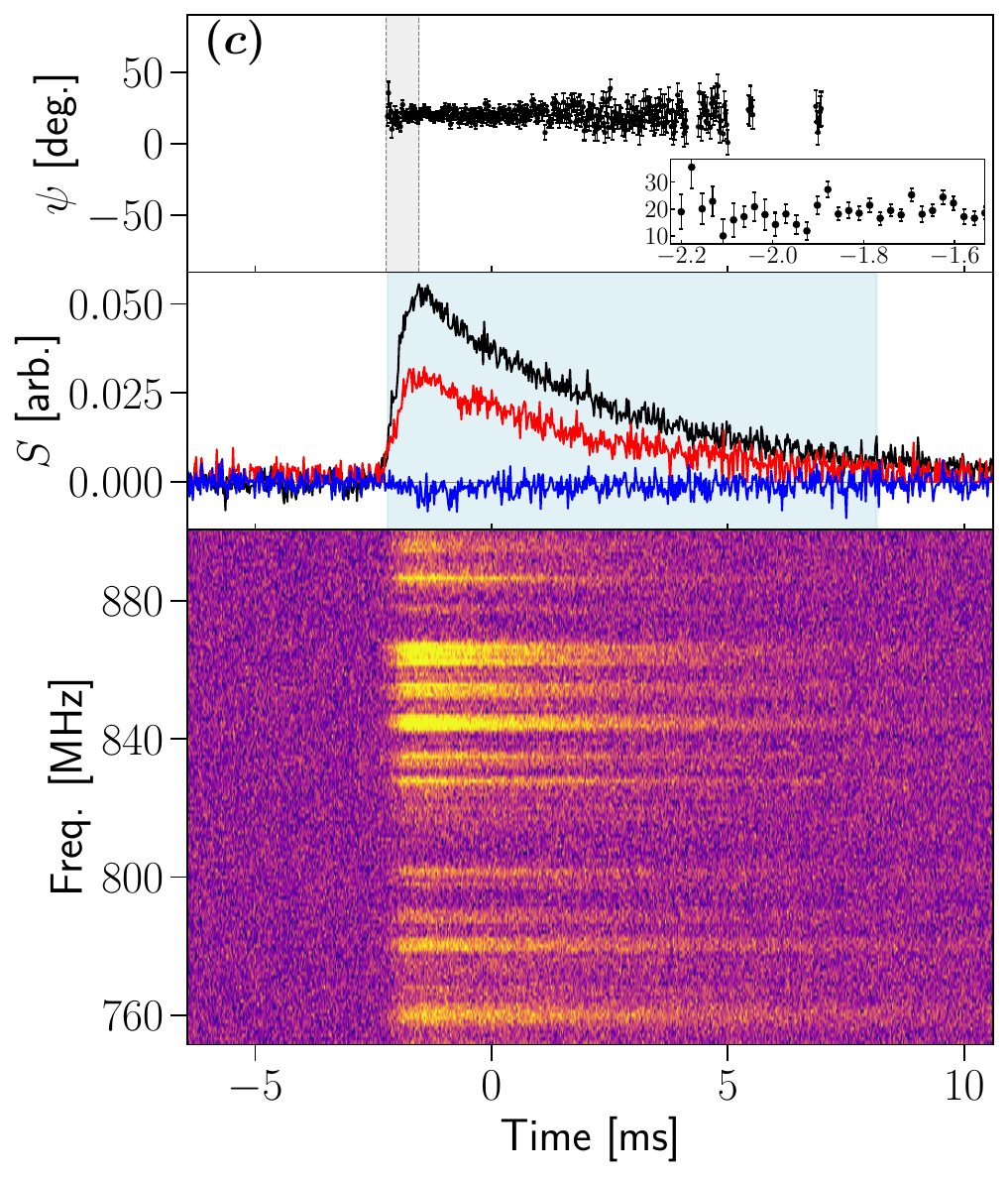}
        \captionlistentry{}\label{fig:191001}
    \end{subfigure}
    \begin{subfigure}[b]{0.45\textwidth}
        \includegraphics[width=\textwidth]{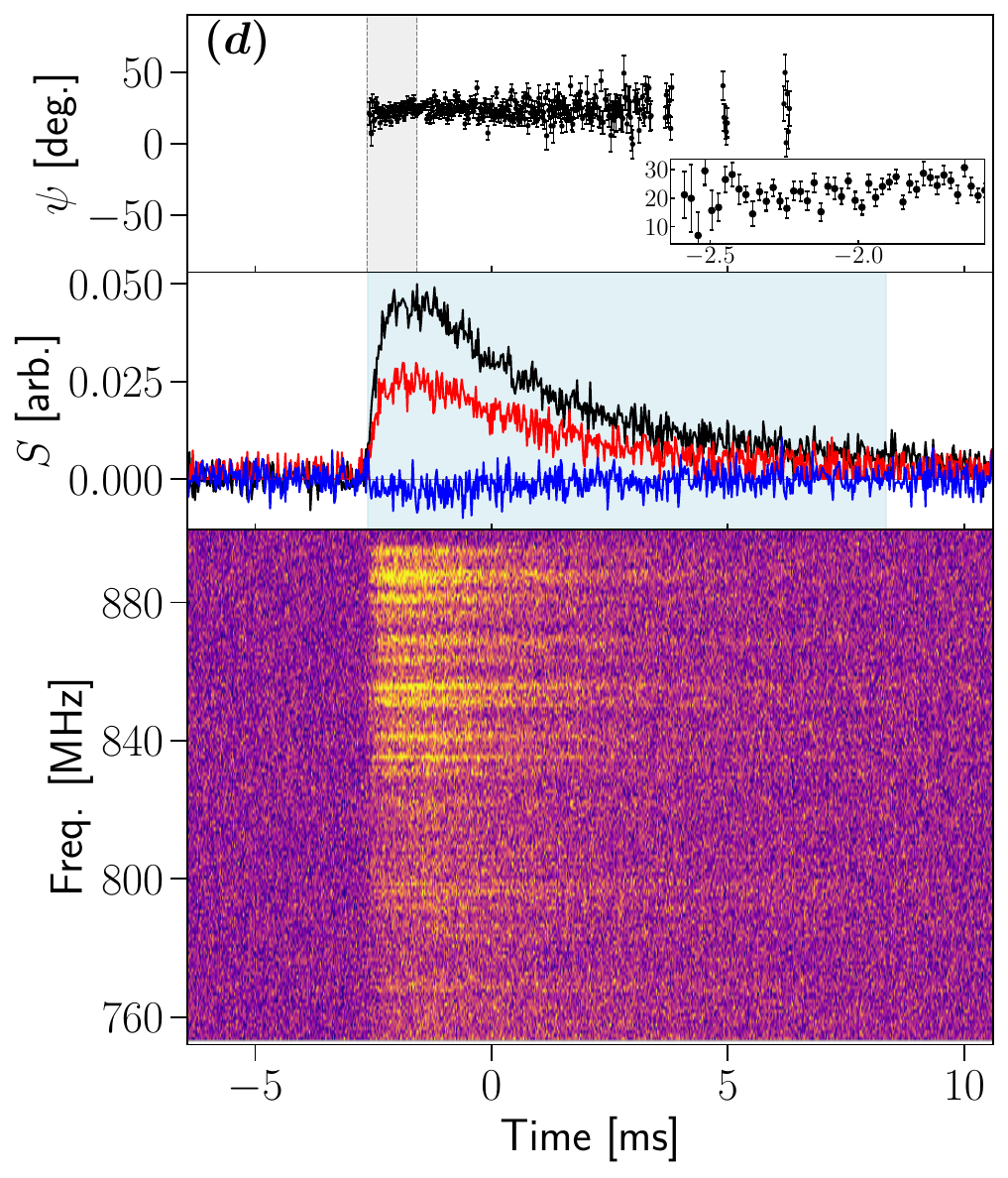}
        \captionlistentry{}\label{fig:HTR 191001}
    \end{subfigure}
    \caption{A \texttt{FIRES} recreation of FRB~20191001A. (a): a noiseless, unscattered FRB comprised of 100 microshots. The bottom panel is the time-frequency dynamic spectrum, the middle panel is the frequency-summed pulse profile (black = total intensity, red = linear polarisation, blue = circular polarisation), and the top panel is the polarisation angle profile. Each top panel includes a zoomed inset spanning the leading phase (from burst onset to the Stokes I peak) to highlight fine PA structure. (b): scattering timescale $\tau_{1\,\rm GHz} = 1.78$\,ms and scintillation added to the top left plot. (c): noise added to the top right plot (on-pulse S/N$=180$). (d): real FRB~20191001A data reproduced from~\cite{2025arXiv250517497S} (on-pulse S/N$=194$, RM corrected from RM=53.47\,rad\,m$^{-2}$). The full list of parameters used are presented in Table~\ref{tab:params} and are described in Section~\ref{sec:model}. The top panels show the polarisation angle, the centre panels show the pulse profile, and the bottom panels show the dynamic spectra. The blue shaded regions in the centre panels is the minimum boxcar width that contains 95\% of the total flux in the pulse profile.}
    \label{fig:FRB_191001_compare}
\end{figure*}

Figure~\ref{fig:FRB_191001_compare} shows the dynamic spectra of three stages of a \texttt{FIRES} simulation. Starting with a noiseless, unscattered FRB comprised of $N=100$ microshots with a uniform distribution of widths ranging from 25--100\,$\mu$s (see Equation~\eqref{eq:width}); the FWHM of the full envelope is 0.5\,ms. Then we add scattering, $\tau_{1\,\rm GHz} = 1.78$\,ms and scintillation decorrelation bandwidth $\nu_s = 2.05$\,MHz with characteristic timescale $t_s=300$\,s, and draw the phase screen with $N_\mathrm{im}=5000$ images and truncation parameter $\theta_\mathrm{lim}=3.0$. Then we add noise ($\mathrm{SEFD}=1.2\rm \,Jy$, see Equation~\eqref{eq:rad}; on-pulse S/N$=213$), and finally compare to real CRAFT FRB~20191001A data reproduced from~\cite{2025arXiv250517497S} with \texttt{ILEX}\footnote{\url{https://github.com/tdial2000/ILEX/tree/main}} (on-pulse S/N$=194$, corrected for RM=53.47 rad\,m$^{-2}$). The simulated dynamic spectra are generated over a frequency range 751.5--900.5\,MHz with 1\,MHz resolution, and a time resolution of 0.023\,ms chosen such that (the real) FRB~20191001A has a peak S/N$\sim20$. The frequency dependence for scattering is $\alpha = -4.85$. Post-processing measurements of the on-pulse polarisation fractions show $\Pi_L = 0.55$ and $\Pi_V = -0.05$. We assume that, intrinsically, each microshot is $\sim 100\%$ polarised, and so we set $\Pi_{L,0} = 0.99$ and $\Pi_{V,0} = -0.05$~\citep[similar to crab nanoshots,][]{2016ApJ...833...47H}. Their PAs are drawn from $\mathcal{N}\left(20.0\,\mathrm{deg}, 30.0\,\mathrm{deg} \right)$. $\tau_{1\,\rm GHz}$ and $\alpha$ are taken from Table 2 of~\cite{2025arXiv250517497S}, while the rest of the parameters are estimates made by-eye to mimic the properties of FRB~20191001A. The top PA panels include zoomed insets of the leading part of the burst to make the unresolved short-timescale PA structure easier to inspect. This initial comparison (Figure~\ref{fig:FRB_191001_compare}) is qualitative and intended to visually demonstrate that \texttt{FIRES} can generate realistic-looking bursts. The remainder of this work focuses on quantitative metrics that enable direct comparison between simulated and real FRBs.
The full list of model parameters are presented in Table~\ref{tab:params}.

\subsection{Polarisation Angle Variance}\label{subsec:PA variance}
FRBs show variable PA properties~\citep[e.g.,][]{2020Natur.586..693L,2025Natur.637...43M}, and when probed at high time resolutions, reveal systematic fluctuations~\citep[][]{2021NatAs...5..594N,2023MNRAS.526.2039H,2025arXiv250517497S}. It has been suggested that the strength and duration of these PA microstructures may correspond to the magnetic field conditions in the emission environment~\citep[e.g.,][]{2025arXiv250517497S}, and so, disentangling intrinsic and extrinsic PA fluctuations is crucial for constraining emission models. Propagation effects such as scattering introduce frequency- and time-dependent effects that obfuscate intrinsic FRB structures, and so, determining under what conditions and which portions of the FRB dynamic spectra still contain original properties is also important.

Figure~\ref{fig:FRB_191001_compare} shows that the PA of the simulated bursts is not constant, but varies with time, scattering timescale, and noise. We define two regions in the on-pulse phase (light-blue-shaded region); the leading region is from the onset of the on-pulse to and including the peak of the burst, and the trailing region is from the peak to the end of the on-pulse. As expected, scattering flattens the PA throughout the trailing region of the burst, reducing intrinsic variability~\citep[][]{2003A&A...410..253L}, while the leading region comparatively remains largely unchanged. The introduction of noise dominates the macrostructure, introducing higher PA variance in the trailing edge. 

To quantify the effects of scattering on PA in different burst regions, we introduce the metric,
\begin{equation}\label{eq:R}
    \mathcal{R}_\psi = \frac{\mathbb{V}(\psi)}{\mathbb{V}(\psi_\mathrm{i})},
\end{equation}
where $\mathbb{V}(\psi)$ is the PA variance across the on-pulse phase and $\mathbb{V}(\psi_\mathrm{i})$ is the PA variance across the individual microshots that form the FRB. This parameter quantifies how much of the intrinsic microshot PA variance remains after taking their superposition. We can approximate the expected (noiseless) PA variance in deg.$^2$ with,
\begin{equation}\label{eq:var exp}
    \mathbb{V}_{\exp}(\psi) \;=\; 
    \frac{\sigma_\psi^2}{N}\!
    \left(\frac{W_{\mathrm{tot}}}{w_{\mathrm{tot}}} - 1\right)\left(\frac{180}{\pi}\right)^2
\end{equation}
where $W_\mathrm{tot}$ and $w_\mathrm{tot}$ are the envelope and microshot FWHMs after scattering. We derive this expression in~\ref{A1}. This approximation is only accurate for cases with high S/N and low $\sigma_{\psi}$. Substituting this into Equation~\eqref{eq:R}, we can find the expected value of $\mathcal{R}_\psi$.

\begin{figure}[ht!]
    \centering

    \begin{subfigure}[t]{\textwidth}
        \centering
        \includegraphics[width=\linewidth]{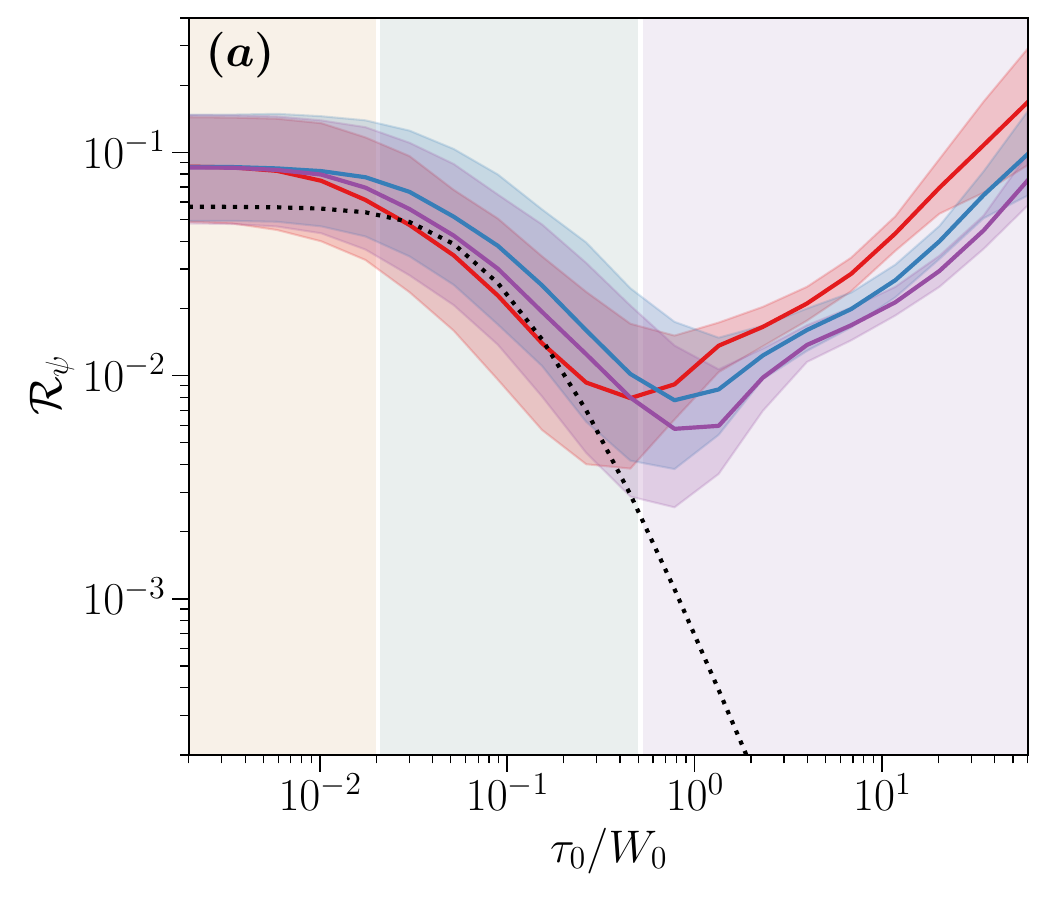}\par\medskip
        \captionlistentry{}\label{fig:PA-quarter}
    \end{subfigure}\vspace{-37pt}\\
    \begin{subfigure}[t]{\textwidth}
        \centering
        \includegraphics[width=\linewidth]{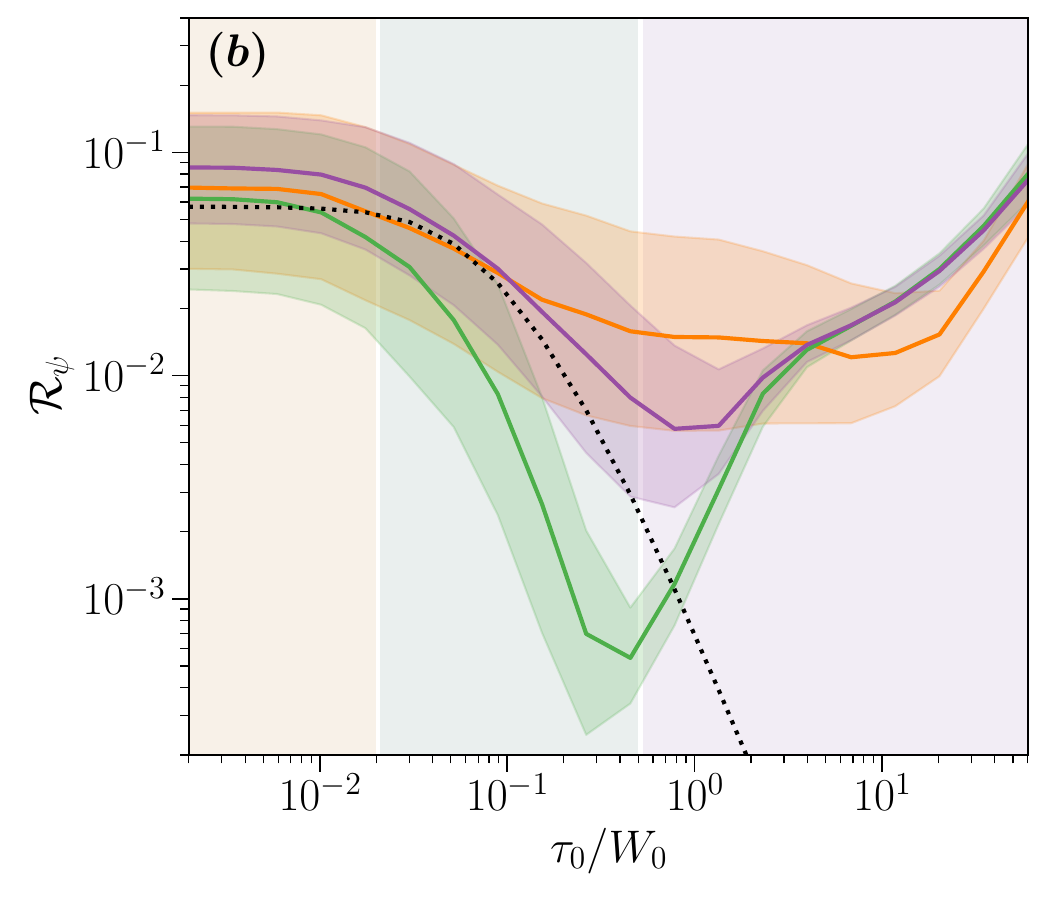}\par\medskip
        \captionlistentry{}\label{fig:PA-phase}
    \end{subfigure}
    \caption{$\mathcal{R}_\psi$ (see Equation~\eqref{eq:R}) versus $\tau_0$ weighted by initial Gaussian envelope width, $W_0$, for a high S/N case of the mock FRB~20191001A (Figure~\ref{fig:191001}). The solid lines are the median value of 500 random FRB realisations at each scattering timescale, and the shaded regions represent the 16th and 84th percentiles of the realisations. The black-dotted line is the (noiseless) expected value from Equation~\eqref{eq:var exp}. $\mathcal{R}_\psi$ exhibits three regimes, highlighted by the light orange, light green, and light purple shaded regions, respectively: a negligible-scattering regime where $\mathcal{R}_\psi$ remains approximately constant, an intermediate regime where scattering blends random microshots together and reduces the variance, and a large-scattering regime where noise dominates and the variance increases. (a): frequency band comparison; The red line is the contribution from the lowest quarter of the band, the blue line is the contribution from the highest quarter of the band, and the purple line is the contribution from the full band. (b): phase comparison; The orange line is the contribution from the first half of the burst, the green line is the contribution from the second half of the burst, and the purple line is the contribution from the entire burst. At $\tau_0/W_0=0, 60$, S/N $\sim 2300, 150$, respectively.}
    \label{fig:PA-combined-3x2}
\end{figure}

In Figure~\ref{fig:PA-combined-3x2} we show how $\mathcal{R}_\psi$ modulates as a function of the input scattering timescale, $\tau_0$, weighted by the input FWHM of the envelope, $W_0$, for different frequency bands and phase regions of a high S/N case of the simulated FRB shown in Figure~\ref{fig:191001}. $\mathcal{R}_\psi$ exhibits three regimes, which are highlighted in the figure by light orange, light green, and light purple shaded regions, respectively. In the negligible-scattering regime, $\mathcal{R}_\psi$ remains constant with $\tau_0$, since the scattering timescale is small compared to the intrinsic PA variation timescales (i.e., the width of each microshot). As $\tau_0$ increases, $\mathcal{R}_\psi$ then decreases, as scattering blends random microshots together, reducing the variance. At very large $\tau_0$, noise begins to dominate and variance increases. As expected, $\mathcal{R}_\psi$ in the lower frequency band is initially lower due to the frequency dependence of scattering, but becomes higher at large $\tau_0$ once the PA is flattened by scattering and noise dominates. The trailing region follows the same trend but is more strongly affected by scattering, leading to a more rapid decline in $\mathcal{R}_\psi$. Similarly, the highest-quarter and full-band with total-phase, also decrease as they become scatter dominated, but transition to the noise-dominated regime at larger $\tau_0$ owing to their higher S/N.

By contrast, the leading edge consistently retains a larger fraction of the intrinsic PA variance at increasing scattering timescales. In high S/N cases, this makes the leading edge the most robust phase region for preserving information about the intrinsic PA structure, and therefore the most informative probe of the initial emission and generation conditions.

The black dotted line shows the expected value of $\mathcal{R}_\psi$ (Equation~\eqref{eq:var exp}) in the noiseless regime and is in general agreement with all regions in the low-scattering, high-S/N limit.

\subsection{Linear Polarisation Fraction}\label{subsec:linpol}
FRBs generally exhibit high linear polarisation fractions~\citep[][]{2024ApJ...964..131S, 2024ApJ...968...50P,2025arXiv250517497S}, and so it has been thought that all FRBs are intrinsically 100\% linearly polarised, with any depolarisation or polarisation conversion occurring due to rotation measure (RM) scattering~\citep[][]{2022Sci...375.1266F,2025arXiv250319749U}. However, \cite{2025ApJ...979..160S} and \cite{2025arXiv250517497S} show no correlation between scattering and polarisation fraction. Further, \cite{2025arXiv250517497S} rule out depolarisation due to unresolved, random PA structures as there is no correlation between decreases in linear polarisation fraction and PA fluctuations. \cite{2024ApJ...964..131S} also suggest that this case may be less significant than propagation effects.

Within the PSN framework, Figure~\ref{fig:FRB_191001_compare} provides an illustrative example in which depolarisation arises from the superposition of microshots with differing PAs. The intrinsic microshot polarisation fractions are $\Pi_{L,0} = 0.99$ and $\Pi_{V,0} = -0.05$; however, when superimposing $N=100$ microshots with PAs sampled from a normal distribution with a standard deviation of $\sigma_\psi=30^\circ$, then $\Pi_{L} = 0.55$ as measured from the real FRB. 

\begin{figure*}[ht!]
    \centering
    \begin{subfigure}{0.45\textwidth}
        \centering
        \includegraphics[
            width=\textwidth,
            height=0.95\textwidth,
            keepaspectratio,
            clip
        ]{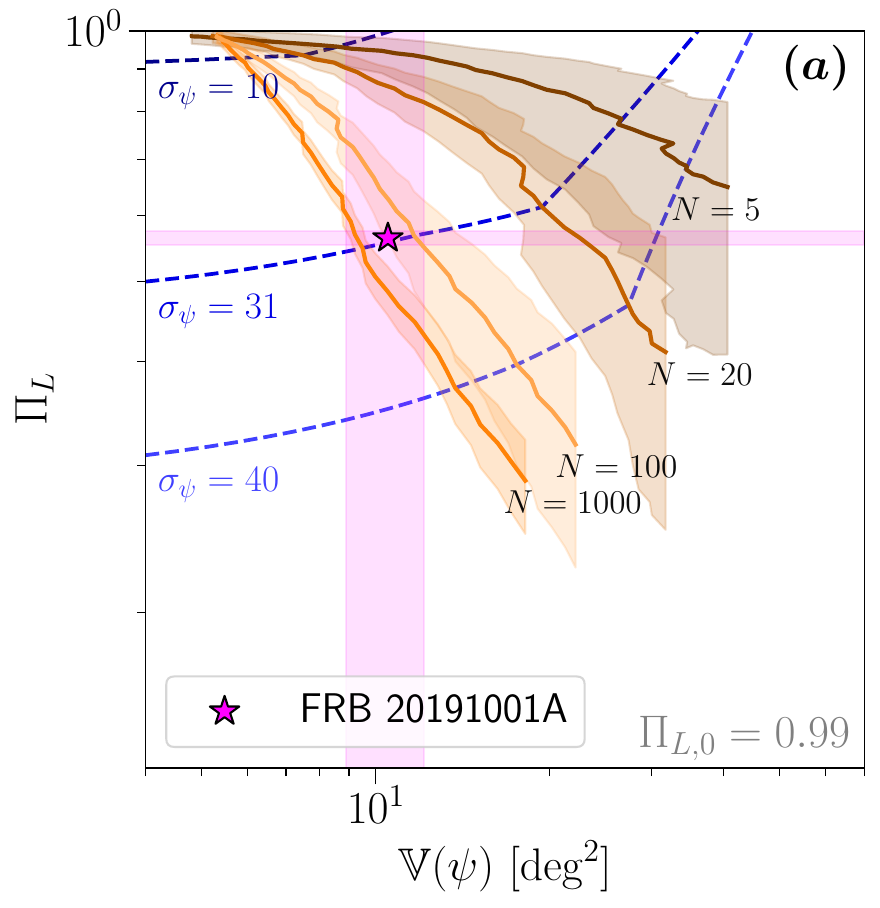}
        \captionlistentry{}\label{fig:Lfrac-99}
    \end{subfigure}\vspace{-31pt}
    \begin{subfigure}{0.45\textwidth}
        \centering
        \includegraphics[
            width=\textwidth,
            height=0.95\textwidth,
            keepaspectratio,
            clip
        ]{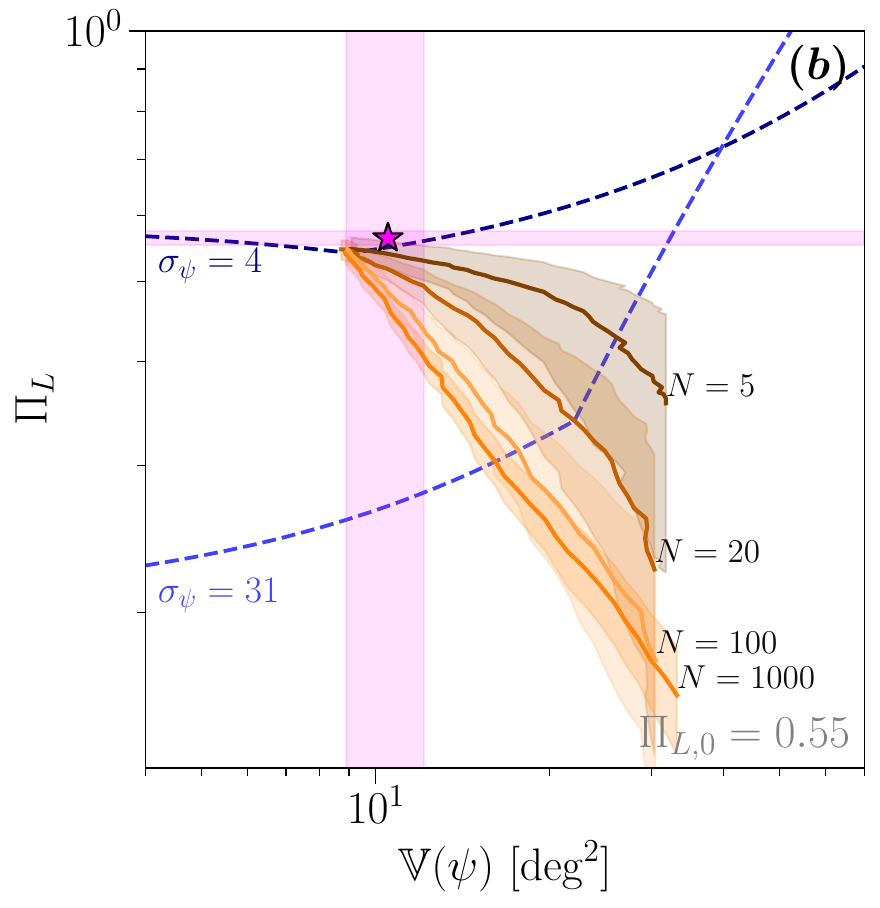}
        \captionlistentry{}\label{fig:Lfrac-55}
    \end{subfigure}\\
    \begin{subfigure}{0.45\textwidth}
        \centering
        \includegraphics[
            width=\textwidth,
            height=0.95\textwidth,
            keepaspectratio,
            clip
        ]{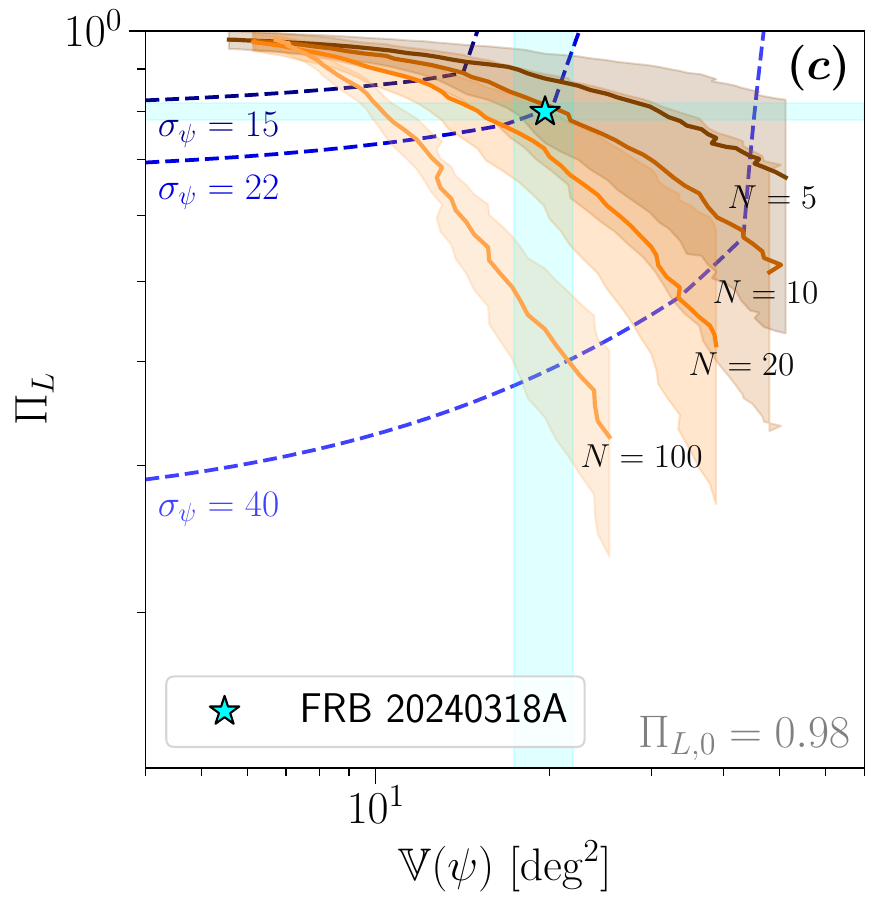}
        \captionlistentry{}\label{fig:240318A-Lfrac-98}
    \end{subfigure}
    \begin{subfigure}{0.45\textwidth}
        \centering
        \includegraphics[
            width=\textwidth,
            height=0.95\textwidth,
            keepaspectratio,
            clip
        ]{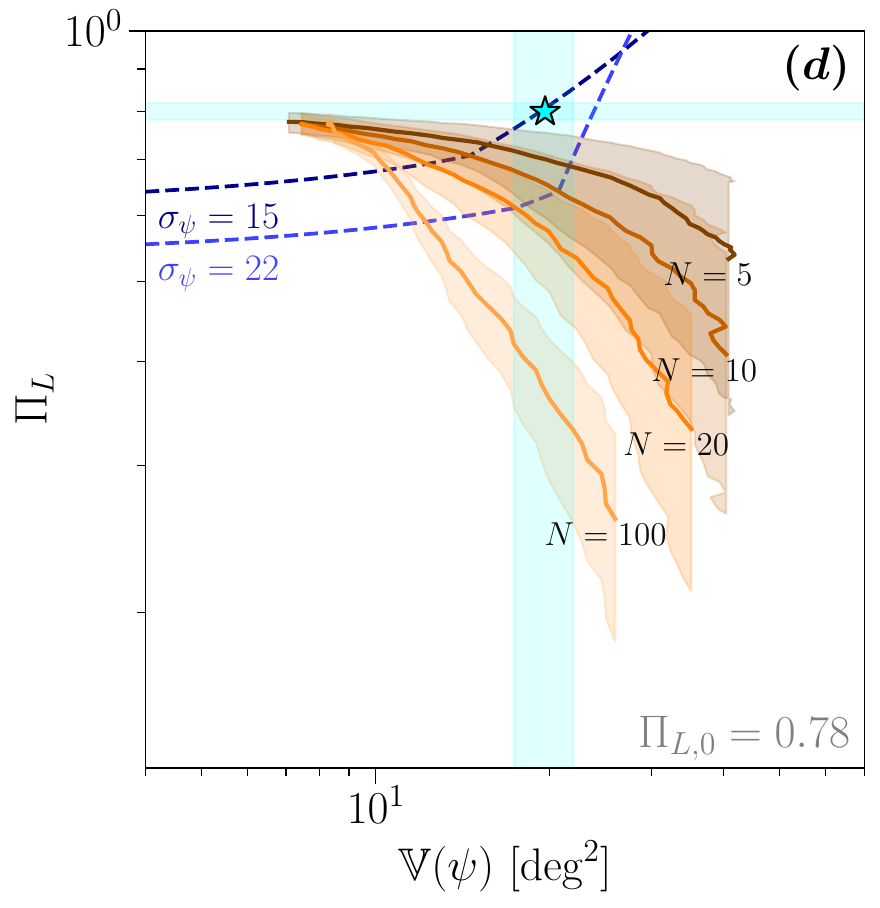}
        \captionlistentry{}\label{fig:240318A-Lfrac-78}
    \end{subfigure}
    
    \caption{Measured linear polarisation fraction, $\Pi_{L}$, versus measured PA variance, $\mathbb{V}(\psi)$, as a function of the standard deviation of the intrinsic microshot PA, $\sigma_\psi$, for the leading phases of the mock FRB~20191001A (panels (a)–(b) with intrinsic linear polarisation fractions $\Pi_{L,0}=0.99$ and $0.55$, respectively) and FRB~20240318A (panels (c)–(d), $\Pi_{L,0}=0.98$ and $red{0.78}$). For $N=5,10,20,100,1000$ microshots, at each $\sigma_\psi$ we generate 500 random realisations of each FRB with fixed $\mathrm{S/N}$ (FRB~20191001A: median $\sim 194$; FRB~20240318A: median $\sim110$) and plot the medians of the $\Pi_{L}$ and $\mathbb{V}(\psi)$ distributions as solid lines. The shaded regions mark only the 16th–84th percentiles of the $\Pi_{L}$ distributions; they do not show percentile ranges in $\mathbb{V}(\psi)$. The magenta and cyan stars show the measured values for FRB~20191001A and FRB~20240318A, respectively (see Figures~\ref{fig:HTR 191001} and \ref{fig:HTR 240318A}), with shaded bands indicating errors from off-pulse RMS noise. Blue dashed lines connect points of constant $\sigma_\psi$ and are linearly extended using the slopes of their first and last segments (FRB~20191001A: $\sigma_\psi = 4^\circ, 10^\circ, 31^\circ, 40^\circ$; FRB~20240318A: $\sigma_\psi = 15^\circ, 22^\circ, 40^\circ$).}
    \label{fig:LV}
\end{figure*}

We illustrate this for the leading edge (see Section~\ref{subsec:PA variance}) of FRB~20191001A in Figures~\ref{fig:LV}\subref{fig:Lfrac-99},\subref{fig:Lfrac-55} where we sweep $\sigma_\psi=0$--$45^\circ$ for different values of $N$. In Figure~\ref{fig:LV}, the shaded percentile regions quantify the spread in $\Pi_L$ only; they do not represent the spread in $\mathbb{V}(\psi)$. Starting with $N=5$ in Figure~\ref{fig:Lfrac-99}, the left-most point corresponds to $\sigma_\psi=0^\circ$, and the right-most point corresponds to $\sigma_\psi=45^\circ$. At $\sigma_\psi=0^\circ$ the intrinsic PA profile is completely flat, no depolarisation due to microshot superposition occurs and so $\Pi_{L} = \Pi_{L,0} = 0.99$. Here the only contribution to $\mathbb{V}(\psi)$ is from noise, as described by Equation~\eqref{eq:rad} which determines the minimum $\mathbb{V}(\psi)$ for all values of $N$ at $\sigma_\psi=0^\circ$. As we increase $\sigma_\psi$, $\mathbb{V}(\psi)$ increases and microshots with different PAs start overlapping, causing a reduction in $\Pi_L$. As we move to larger values of $N$, we increase the effective shot-rate ($N_{\rm eff}\,{\rm shots\,}s^{-1}$; see~\ref{A1}) since the envelope width, $W_0$, remains constant. So, as $N$ increases, microshots overlap more often and stronger depolarisation occurs for a given $\sigma_\psi>0$. This also means that, since more averaging of PAs occurs, $\mathbb{V}(\psi)$ decreases for a given $\sigma_\psi>0$. If we decrease $\Pi_{L,0}$, we decrease our signal in $L$ and noise becomes more significant. So a decrease in $\Pi_{L,0}$ also corresponds to an increase in $\mathbb{V}(\psi)$ which we see when comparing to Figure~\ref{fig:Lfrac-55} where we have reduced $\Pi_{L,0}$ to 0.55. Under the PSN model, we reproduce observable parameters and use these tracks to identify allowed combinations of $N$ and $\sigma_\psi$ for assumed $\Pi_{L,0}$, rather than to uniquely constrain a solution. The allowed regions are therefore conditional on the adopted microshot width and amplitude distributions (\ref{subsec: amp dist}, \ref{subsec: width dist}).

We show the same analysis for FRB~20240318A in Figures~\ref{fig:LV}\subref{fig:240318A-Lfrac-98},\subref{fig:240318A-Lfrac-78}
and~\ref{fig:FRB_240318A_compare} using the parameters listed in Table~\ref{tab:params}. Post-processing measurements of the on-pulse polarisation fractions in Figure~\ref{fig:FRB_240318A_compare}\subref{fig:HTR 240318A} show $\Pi_L = 0.78$ and $\Pi_V = -0.17$.

\section{DISCUSSION}\label{sec: discussion}
Here, we first show that \texttt{FIRES} qualitatively reproduces FRB behaviour, then proceed to quantitatively constrain the allowed parameter space. In order to visually reproduce the properties of FRB~20191001A (Figure~\ref{fig:191001}), we adopt a ``single-component'' envelope (Table~2 of \citealt{2025arXiv250517497S}), implying an intrinsic FWHM $W_0 \sim 0.5$\,ms given the measured scattering timescale $\tau_{\rm 1\,GHz}=1.78$\,ms. As an illustrative example within the PSN model, we choose $N=100$ microshots with intrinsic linear polarisation fraction $\Pi_{L,0}=0.99$ arriving within $W_0$, for which matching the observed linear polarisation fraction requires $\sigma_\psi \sim 30^\circ$ (Figure~\ref{fig:LV}\subref{fig:Lfrac-99}). In this example, superposition of microshots with differing intrinsic PAs produces a time-variable PA profile, while scattering preferentially suppresses this variability on the trailing edge, broadly consistent with observations \citep{2025arXiv250517497S}. While more rigorous statistical comparisons against the full HTR CRAFT sample remain for future work, the quantitative constraints below demonstrate the viability of the PSN framework.

\subsection{Depolarisation Mechanisms and Parameter Degeneracy}\label{subsec: depol degen}
Scattering is not the sole mechanism that suppresses observable PA structure. Finite instrumental sampling further averages intrinsic PA fluctuations~\citep[][]{2022MNRAS.510.4654B}: when the sampling time $\delta t$ exceeds the characteristic microshot separation, rapid PA variations are smoothed and depolarisation occurs, yielding an apparently flat PA profile even for relatively large $\sigma_\psi$. A similar degeneracy exists in the construction of the burst envelope itself: in our simulations the envelope is produced by varying the number of microshots with amplitudes, $A_i$, sampled from a power law and uniformly sampled widths, $w_i$, but comparable averaged behaviour could also arise from a constant number of shots with time-varying amplitudes, and different amplitude distributions and ranges, an effect that would likewise be smoothed by superposition and finite sampling (see~\ref{subsec: amp dist} and~\ref{subsec: width dist}).

In the PSN framework, depolarisation is set mainly by effective averaging: microshot overlap ($N\,w_i/W_0$), propagation smearing, and instrumental sampling. For $N\,w_i \gg W_0$, vector averaging in $(Q,U)$ suppresses $\Pi_L$ and flattens or disorders PA; for $N\,w_i \lesssim W_0$, shots are more resolved, so strong PA variation can coexist with comparatively high $\Pi_L$.

The Appendix and simulations show the same behaviour in $\Delta\psi \equiv \psi-\bar{\psi}$ versus $\Delta\Pi_{L} \equiv \Pi_{L}-\bar{\Pi}_{L}$ (Figure~\ref{fig:pa-li-full}; ~\ref{sec: PA L/I}): the noiseless no-scattering case is a near-random walk about $(0,0)$, while a clear anti-correlation appears only when scattering is strong enough to enforce substantial averaging. With realistic noise, this anti-correlation is largely erased.

This interpretation is consistent with observations: more highly scattered bursts tend to show flatter PA \citep{2025arXiv250517497S}, yet PA fluctuations do not robustly track $L/I$ \citep{2025arXiv250517497S}, and $L/I$ shows no clear dependence on scattering timescale \citep{2025ApJ...979..160S}. Within PSN, these results imply that large-$\sigma_\psi$ averaging is not the sole or primary depolarisation driver; instead, the observed behaviour reflects a coupled degeneracy among $\sigma_\psi$, $N$, $\tau$, $\delta t$, $A_i$, $w_i$, and S/N, with a likely non-negligible contribution from external depolarising media along the propagation path \citep[e.g.,][]{2026MNRAS.545f1997U}.

\subsection{FRB Case Studies}
These degeneracies are illustrated in Figure~\ref{fig:Lfrac-55}, where all simulations converge to identical $\Pi_L$ and $\mathbb{V}(\psi)$ at $\sigma_\psi=0$. Shifts in $\mathbb{V}(\psi)$ require changes in the effective S/N: increasing S/N reduces the measured PA variance, while decreasing the S/N of the linearly polarised intensity $L$ (lowering $\Pi_{L,0}$) increases the contribution of noise to $\mathbb{V}(\psi)$. The measured values for FRB~20191001A (Figure~\ref{fig:HTR 191001}) therefore do not uniquely select a single intrinsic configuration, nor do they tightly determine the intrinsic linear fraction itself; rather, they define a model-dependent allowed region of $(N,\sigma_\psi,\mathbb{V}(\psi),\Pi_{L,0})$ space.

Under the PSN model with a set of assumptions about microshot width and amplitude distributions, Figure~\ref{fig:Lfrac-99} shows that reproducing the observed properties of FRB~20191001A with intrinsically 100\% polarised microshots ($\Pi_{L,0}=0.99$, $\Pi_{V,0}=-0.05$) is compatible with intrinsic PA dispersions of order $\sigma_\psi \sim 31^\circ$ when the number of microshots is $N \gtrsim 100$. This reflects efficient averaging of intrinsically misaligned PAs through microshot superposition, scattering, and finite instrumental sampling. However, once these averaging effects and the contribution of noise are accounted for, comparable levels of measured PA variance can also be obtained without invoking such large intrinsic dispersions. In particular, Figure~\ref{fig:Lfrac-55} demonstrates that reducing the intrinsic linear fraction to the measured value of $\Pi_{L,0}=0.55$ allows models with more modest PA dispersion ($\sigma_\psi \sim 4^\circ$) and $N \lesssim 5$ to reproduce the observations. Larger values of $N$ combined with slightly reduced $\Pi_{L,0}$ and smaller $\sigma_\psi$ remain equally consistent. Taken together, the data for FRB~20191001A are best described by a broad, model-dependent region of parameter space in which modest intrinsic PA dispersion is compatible with the observed near-constant PA once instrumental resolution, propagation effects, noise statistics, and the assumed microshot-property distributions are properly incorporated.

By contrast, FRB~20240318A exhibits a significant, observable variable PA trend \citep{2025arXiv250517497S}. Assuming a 100\% polarised microshots ($\Pi_{L,0}=0.98$, $\Pi_{V,0}=-0.17$), Figure~\ref{fig:240318A-Lfrac-98} allows configurations with $N=5$--20 and intrinsic PA dispersion $\sigma_\psi \sim 22^\circ$. Relaxing this assumption to the measured $\Pi_{L,0}=0.78$ shifts the allowed region toward fewer microshots, with Figure~\ref{fig:240318A-Lfrac-78} favouring $N < 5$ and more modest PA dispersion ($\sigma_\psi \sim 15^\circ$). A stricter interpretation of the resolved PA structure is that the data contain multiple independent PA excursions across the burst window; under that interpretation, very low-$N$ solutions (e.g., $N<5$) are less physically plausible in the PSN framework unless each observable excursion is produced by unresolved blends of multiple microshots. These solutions therefore occupy a different part of the same $(N,\sigma_\psi,\Pi_{L,0})$ space explored for FRB~20191001A, rather than requiring a fundamentally larger intrinsic PA dispersion. The observed PA variability therefore reflects a combination of reduced microshot averaging and intrinsic PA structure, rather than uniquely implying broader intrinsic PA distributions.

\subsection{Model Alternatives and Interpretation}
At the phenomenological level considered here, the main alternative to a multi-shot description is a single-shot or effectively single-component model in which the observed burst structure is generated within one emission episode rather than by superposition of many narrower shots. We do not rule out such alternatives in the present work. However, a single-shot picture appears to struggle more naturally to explain the rapid, $\mu$s PA variations seen in HTR bursts~\citep[e.g.,][]{2022NatAs...6..393N,2025arXiv250517497S}, whereas the PSN framework can reproduce qualitatively and quantitatively similar behaviour through the superposition, propagation, and sampling of many polarised microshots. We therefore view the current results as providing a first-order demonstration that the PSN model is viable over a broad parameter space, rather than as definitive evidence that it is uniquely required. Stronger support for, or against, the shot-noise interpretation will require a future statistical comparison of \texttt{FIRES} against the growing sample of HTR CRAFT FRBs and, ideally, against explicitly formulated alternative models.

\subsection{Limitations and Caveats}
A key limitation of the present analysis is the use of PA variance as the primary diagnostic of intrinsic structure. Variance is not sensitive to how the PA varies in time, encoding neither coherence, intermittency, nor ordered swings. Consequently, distinct PA behaviours can yield similar $\mathbb{V}(\psi)$, particularly in noise-dominated or temporally averaged regimes (see Figure~\ref{fig:PA-combined-3x2}). Likewise, the present implementation treats the intrinsic PA structure primarily as stochastic microshot-to-microshot scatter, and so does not yet capture more complicated time-dependent PA evolution, such as coherent swings or other structured behaviour across the burst envelope. Structure-based metrics that explicitly quantify PA organisation \citep[e.g.,][]{2025arXiv250517497S} are therefore likely to provide a more informative and robust characterisation of intrinsic PA behaviour than variance alone. Future extensions of \texttt{FIRES} will incorporate such measures.

Finally, comparisons between FRB microstructure and Galactic pulsars such as the Crab must be interpreted cautiously. The extreme PA swings observed in Crab nanoshots occur on nanosecond timescales and would be strongly averaged out at the millisecond-to-microsecond resolutions of current FRB observations. As a result, present data probe only a temporally averaged manifestation of any underlying nanoshot structure, and unresolved Crab-like PA variability in FRBs cannot be ruled out. In addition, for very bright pulses observed at high time resolution, the inferred polarimetric variability can itself be affected by self-noise and small-number statistics when the averaging is insufficient \citep{2009ApJ...694.1413V}. Apparent differences in polarimetric behaviour may therefore reflect observational limitations rather than fundamentally distinct emission physics.

\subsection{Future Model Considerations}
The present iteration of \texttt{FIRES} lacks several physical ingredients that will be incorporated into future versions:

\begin{itemize}
    \item Multiple scattering and scintillation screens: Currently only a single thin-screen model is implemented, preventing realistic modelling of multi-path propagation and frequency-dependent scintillation structure.
    \item Thick-screen scattering: The code assumes an infinitely thin screen, whereas a volumetric or extended medium would alter temporal broadening and pulse-shape evolution.
    \item Generalised Faraday rotation: \texttt{FIRES} uses standard Faraday rotation, but cannot yet capture mode-coupling or birefringent effects expected in magnetised, relativistic plasmas.
    \item Structured intrinsic PA evolution and PA diagnostics: Future versions should incorporate more complicated time-dependent PA evolution, including coherent swings across the burst envelope, together with structure-based diagnostics that quantify PA organisation rather than variance alone.
    \item Alternative noise models: For the ASKAP data considered here, Gaussian noise is adequate, but at substantially higher time and frequency resolution the noise can become non-Gaussian. Future applications of \texttt{FIRES} should therefore generate the FRB and noise as electric fields at Nyquist resolution, allowing for noise to occupy a $\chi^2$ distribution.
    \item Intrinsic spectral structure: Extending \texttt{FIRES} to include intrinsic spectral structure would improve the interpretation of its effects on FRB polarisation and flux variability.
\end{itemize}

\section{Conclusions}\label{sec: conclusions}
We have introduced \texttt{FIRES}, an emission-mechanism-independent framework for modelling FRB dynamic spectra as the superposition of Gaussian microshots with varying polarisation angles, followed by propagation effects and noise. Applied to CRAFT FRBs~20191001A and~20240318A, \texttt{FIRES} reproduces key polarimetric behaviours: scattering suppresses PA variability on burst trailing edges, while the leading edge preferentially preserves intrinsic structure. Depolarisation arises naturally from incoherent microshot superposition, with the relative contributions of $N$, $\sigma_\psi$, $\mathbb{V}(\psi)$, and $\Pi_{L,0}$ defining a degenerate but physically bounded parameter space.

For both FRBs, modest intrinsic PA dispersion remains viable once scattering, sampling, and noise are properly accounted for, and current data do not require fundamentally distinct intrinsic PA statistics between the sources. The leading-edge window and higher-frequency sub-bands provide the most robust access to intrinsic information. However, PA variance alone is an incomplete statistic and does not capture how PA varies in time; more structure-sensitive diagnostics are needed to fully characterise intrinsic PA evolution.

Future work will extend \texttt{FIRES} to incorporate multiple scattering screens, generalised Faraday rotation, and intrinsic spectral structure. Observationally, application to high-time-resolution, low-scattering FRBs will test whether \texttt{FIRES} can reliably recover intrinsic microshot properties and break current parameter degeneracies. Such observations will also directly test whether FRBs exhibit Crab-like nanoshot PA variability on unresolved timescales.

\begin{acknowledgement}
This research was supported by an Australian Government Research Training Program (RTP) Scholarship (\url{https://doi.org/10.82133/C42F-K220}). AB acknowledges support through project CORTEX (NWA.1160.18.316) of the research programme NWA-ORC which is financed by the Dutch Research Council (NWO). 
\end{acknowledgement}

\section*{Data Availability Statement}
ASKAP data used in this work are available online at \url{https://doi.org/10.25917/1RG2-C612}. The \texttt{FIRES} codebase and simulated data are available at \url{https://github.com/JoelBalzan/FIRES}.

\bibliography{bibliography}

@ARTICLE{Lorimer2007,
       author = {{Lorimer}, D.~R. and {Bailes}, M. and {McLaughlin}, M.~A. and {Narkevic}, D.~J. and {Crawford}, F.},
        title = "{A Bright Millisecond Radio Burst of Extragalactic Origin}",
      journal = {Science},
         year = 2007,
        month = nov,
       volume = {318},
       number = {5851},
        pages = {777},
          doi = {10.1126/science.1147532},
archivePrefix = {arXiv},
       eprint = {0709.4301},
 primaryClass = {astro-ph},
       adsurl = {https://ui.adsabs.harvard.edu/abs/2007Sci...318..777L}
}

@ARTICLE{2019A&ARv..27....4P,
       author = {{Petroff}, E. and {Hessels}, J.~W.~T. and {Lorimer}, D.~R.},
        title = "{Fast radio bursts}",
      journal = {\aapr},
         year = 2019,
        month = dec,
       volume = {27},
       number = {1},
          eid = {4},
        pages = {4},
          doi = {10.1007/s00159-019-0116-6},
archivePrefix = {arXiv},
       eprint = {1904.07947},
 primaryClass = {astro-ph.HE},
       adsurl = {https://ui.adsabs.harvard.edu/abs/2019A&ARv..27....4P}
}

@ARTICLE{2022A&ARv..30....2P,
       author = {{Petroff}, E. and {Hessels}, J.~W.~T. and {Lorimer}, D.~R.},
        title = "{Fast radio bursts at the dawn of the 2020s}",
      journal = {\aapr},
         year = 2022,
        month = dec,
       volume = {30},
       number = {1},
          eid = {2},
        pages = {2},
          doi = {10.1007/s00159-022-00139-w},
archivePrefix = {arXiv},
       eprint = {2107.10113},
 primaryClass = {astro-ph.HE},
       adsurl = {https://ui.adsabs.harvard.edu/abs/2022A&ARv..30....2P}
}

@ARTICLE{2025arXiv250517497S,
       author = {{Scott}, D.~R. and {Dial}, T. and {Bera}, A. and {Deller}, A.~T. and {Glowacki}, M. and {Gourdji}, K. and {James}, C.~W. and {Shannon}, R.~M. and {Bannister}, K.~W. and {Ekers}, R.~D. and {Sammons}, M. and {Sutinjo}, A.~T. and {Uttarkar}, P.~A.},
        title = "{High-time-resolution properties of 35 fast radio bursts detected by the Commensal Real-time ASKAP Fast Transients Survey}",
      journal = {arXiv e-prints},
         year = 2025,
        month = may,
          eid = {arXiv:2505.17497},
        pages = {arXiv:2505.17497},
          doi = {10.48550/arXiv.2505.17497},
archivePrefix = {arXiv},
       eprint = {2505.17497},
 primaryClass = {astro-ph.HE},
       adsurl = {https://ui.adsabs.harvard.edu/abs/2025arXiv250517497S}
}

@ARTICLE{2007ApJ...670..693H,
       author = {{Hankins}, T.~H. and {Eilek}, J.~A.},
        title = "{Radio Emission Signatures in the Crab Pulsar}",
      journal = {\apj},
         year = 2007,
        month = nov,
       volume = {670},
       number = {1},
        pages = {693-701},
          doi = {10.1086/522362},
archivePrefix = {arXiv},
       eprint = {0708.2505},
 primaryClass = {astro-ph},
       adsurl = {https://ui.adsabs.harvard.edu/abs/2007ApJ...670..693H}
}

@ARTICLE{2016ApJ...833...47H,
       author = {{Hankins}, T.~H. and {Eilek}, J.~A. and {Jones}, G.},
        title = "{The Crab Pulsar at Centimeter Wavelengths. II. Single Pulses}",
      journal = {\apj},
         year = 2016,
        month = dec,
       volume = {833},
       number = {1},
          eid = {47},
        pages = {47},
          doi = {10.3847/1538-4357/833/1/47},
archivePrefix = {arXiv},
       eprint = {1608.08881},
 primaryClass = {astro-ph.HE},
       adsurl = {https://ui.adsabs.harvard.edu/abs/2016ApJ...833...47H}
}

@ARTICLE{2022MNRAS.514.5866G,
       author = {{Gupta}, V. and {Flynn}, C. and {Farah}, W. and {Bailes}, M. and {Deller}, A.~T. and {Day}, C.~K. and {Lower}, M.~E.},
        title = "{The ultranarrow FRB20191107B, and the origins of FRB scattering}",
      journal = {\mnras},
         year = 2022,
        month = aug,
       volume = {514},
       number = {4},
        pages = {5866-5878},
          doi = {10.1093/mnras/stac1720},
archivePrefix = {arXiv},
       eprint = {2209.00311},
 primaryClass = {astro-ph.HE},
       adsurl = {https://ui.adsabs.harvard.edu/abs/2022MNRAS.514.5866G}
}

@ARTICLE{2023NatAs...7.1486S,
       author = {{Snelders}, M.~P. and {Nimmo}, K. and {Hessels}, J.~W.~T. and {Bensellam}, Z. and {Zwaan}, L.~P. and {Chawla}, P. and {Ould-Boukattine}, O.~S. and {Kirsten}, F. and {Faber}, J.~T. and {Gajjar}, V.},
        title = "{Detection of ultra-fast radio bursts from FRB 20121102A}",
      journal = {Nature Astronomy},
         year = 2023,
        month = dec,
       volume = {7},
        pages = {1486-1496},
          doi = {10.1038/s41550-023-02101-x},
archivePrefix = {arXiv},
       eprint = {2307.02303},
 primaryClass = {astro-ph.HE},
       adsurl = {https://ui.adsabs.harvard.edu/abs/2023NatAs...7.1486S}
}

@ARTICLE{2020MNRAS.498..651B,
       author = {{Beniamini}, Paz and {Kumar}, Pawan},
        title = "{What does FRB light-curve variability tell us about the emission mechanism?}",
      journal = {\mnras},
         year = 2020,
        month = oct,
       volume = {498},
       number = {1},
        pages = {651-664},
          doi = {10.1093/mnras/staa2489},
archivePrefix = {arXiv},
       eprint = {2007.07265},
 primaryClass = {astro-ph.HE},
       adsurl = {https://ui.adsabs.harvard.edu/abs/2020MNRAS.498..651B}
}

@ARTICLE{2022NatAs...6..393N,
       author = {{Nimmo}, K. and {Hessels}, J.~W.~T. and {Kirsten}, F. and {Keimpema}, A. and {Cordes}, J.~M. and {Snelders}, M.~P. and {Hewitt}, D.~M. and {Karuppusamy}, R. and {Archibald}, A.~M. and {Bezrukovs}, V. and {Bhardwaj}, M. and {Blaauw}, R. and {Buttaccio}, S.~T. and {Cassanelli}, T. and {Conway}, J.~E. and {Corongiu}, A. and {Feiler}, R. and {Fonseca}, E. and {Forss{\'e}n}, O. and {Gawro{\'n}ski}, M. and {Giroletti}, M. and {Kharinov}, M.~A. and {Leung}, C. and {Lindqvist}, M. and {Maccaferri}, G. and {Marcote}, B. and {Masui}, K.~W. and {Mckinven}, R. and {Melnikov}, A. and {Michilli}, D. and {Mikhailov}, A.~G. and {Ng}, C. and {Orbidans}, A. and {Ould-Boukattine}, O.~S. and {Paragi}, Z. and {Pearlman}, A.~B. and {Petroff}, E. and {Rahman}, M. and {Scholz}, P. and {Shin}, K. and {Smith}, K.~M. and {Stairs}, I.~H. and {Surcis}, G. and {Tendulkar}, S.~P. and {Vlemmings}, W. and {Wang}, N. and {Yang}, J. and {Yuan}, J.~P.},
        title = "{Burst timescales and luminosities as links between young pulsars and fast radio bursts}",
      journal = {Nature Astronomy},
         year = 2022,
        month = feb,
       volume = {6},
        pages = {393-401},
          doi = {10.1038/s41550-021-01569-9},
archivePrefix = {arXiv},
       eprint = {2105.11446},
 primaryClass = {astro-ph.HE},
       adsurl = {https://ui.adsabs.harvard.edu/abs/2022NatAs...6..393N}
}

@ARTICLE{2025Natur.637...43M,
       author = {{Mckinven}, Ryan and {Bhardwaj}, Mohit and {Eftekhari}, Tarraneh and {Kilpatrick}, Charles D. and {Kirichenko}, Aida and {Pal}, Arpan and {Cook}, Amanda M. and {Gaensler}, B.~M. and {Giri}, Utkarsh and {Kaspi}, Victoria M. and {Michilli}, Daniele and {Nimmo}, Kenzie and {Pearlman}, Aaron B. and {Pleunis}, Ziggy and {Sand}, Ketan R. and {Stairs}, Ingrid and {Andersen}, Bridget C. and {Andrew}, Shion and {Bandura}, Kevin and {Brar}, Charanjot and {Cassanelli}, Tomas and {Chatterjee}, Shami and {Curtin}, Alice P. and {Dong}, Fengqiu Adam and {Eadie}, Gwendolyn and {Fonseca}, Emmanuel and {Ibik}, Adaeze L. and {Kaczmarek}, Jane F. and {Kharel}, Bikash and {Lazda}, Mattias and {Leung}, Calvin and {Li}, Dongzi and {Main}, Robert and {Masui}, Kiyoshi W. and {Mena-Parra}, Juan and {Ng}, Cherry and {Pandhi}, Ayush and {Patil}, Swarali Shivraj and {Prochaska}, J. Xavier and {Rafiei-Ravandi}, Masoud and {Scholz}, Paul and {Shah}, Vishwangi and {Shin}, Kaitlyn and {Smith}, Kendrick},
        title = "{A pulsar-like polarization angle swing from a nearby fast radio burst}",
      journal = {\nat},
         year = 2025,
        month = jan,
       volume = {637},
       number = {8044},
        pages = {43-47},
          doi = {10.1038/s41586-024-08184-4},
       adsurl = {https://ui.adsabs.harvard.edu/abs/2025Natur.637...43M}
}

@ARTICLE{2020Natur.586..693L,
       author = {{Luo}, R. and {Wang}, B.~J. and {Men}, Y.~P. and {Zhang}, C.~F. and {Jiang}, J.~C. and {Xu}, H. and {Wang}, W.~Y. and {Lee}, K.~J. and {Han}, J.~L. and {Zhang}, B. and {Caballero}, R.~N. and {Chen}, M.~Z. and {Chen}, X.~L. and {Gan}, H.~Q. and {Guo}, Y.~J. and {Hao}, L.~F. and {Huang}, Y.~X. and {Jiang}, P. and {Li}, H. and {Li}, J. and {Li}, Z.~X. and {Luo}, J.~T. and {Pan}, J. and {Pei}, X. and {Qian}, L. and {Sun}, J.~H. and {Wang}, M. and {Wang}, N. and {Wen}, Z.~G. and {Xu}, R.~X. and {Xu}, Y.~H. and {Yan}, J. and {Yan}, W.~M. and {Yu}, D.~J. and {Yuan}, J.~P. and {Zhang}, S.~B. and {Zhu}, Y.},
        title = "{Diverse polarization angle swings from a repeating fast radio burst source}",
      journal = {\nat},
         year = 2020,
        month = oct,
       volume = {586},
       number = {7831},
        pages = {693-696},
          doi = {10.1038/s41586-020-2827-2},
archivePrefix = {arXiv},
       eprint = {2011.00171},
 primaryClass = {astro-ph.HE},
       adsurl = {https://ui.adsabs.harvard.edu/abs/2020Natur.586..693L}
}

@ARTICLE{2024ApJ...964..131S,
       author = {{Sherman}, Myles B. and {Connor}, Liam and {Ravi}, Vikram and {Law}, Casey and {Chen}, Ge and {Catha}, Morgan and {Faber}, Jakob T. and {Hallinan}, Gregg and {Harnach}, Charlie and {Hellbourg}, Greg and {Hobbs}, Rick and {Hodge}, David and {Hodges}, Mark and {Lamb}, James W. and {Rasmussen}, Paul and {Sharma}, Kritti and {Shi}, Jun and {Simard}, Dana and {Somalwar}, Jean and {Squillace}, Reynier and {Weinreb}, Sander and {Woody}, David P. and {Yadlapalli}, Nitika and {The Deep Synoptic Array team}},
        title = "{Deep Synoptic Array Science: Polarimetry of 25 New Fast Radio Bursts Provides Insights into Their Origins}",
      journal = {\apj},
         year = 2024,
        month = apr,
       volume = {964},
       number = {2},
          eid = {131},
        pages = {131},
          doi = {10.3847/1538-4357/ad275e},
archivePrefix = {arXiv},
       eprint = {2308.06813},
 primaryClass = {astro-ph.HE},
       adsurl = {https://ui.adsabs.harvard.edu/abs/2024ApJ...964..131S}
}

@ARTICLE{2024ApJ...968...50P,
       author = {{Pandhi}, Ayush and {Pleunis}, Ziggy and {Mckinven}, Ryan and {Gaensler}, B.~M. and {Su}, Jianing and {Ng}, Cherry and {Bhardwaj}, Mohit and {Brar}, Charanjot and {Cassanelli}, Tomas and {Cook}, Amanda and {Curtin}, Alice P. and {Kaspi}, Victoria M. and {Lazda}, Mattias and {Leung}, Calvin and {Li}, Dongzi and {Masui}, Kiyoshi W. and {Michilli}, Daniele and {Nimmo}, Kenzie and {Pearlman}, Aaron B. and {Petroff}, Emily and {Rafiei-Ravandi}, Masoud and {Sand}, Ketan R. and {Scholz}, Paul and {Shin}, Kaitlyn and {Smith}, Kendrick and {Stairs}, Ingrid},
        title = "{Polarization Properties of 128 Nonrepeating Fast Radio Bursts from the First CHIME/FRB Baseband Catalog}",
      journal = {\apj},
         year = 2024,
        month = jun,
       volume = {968},
       number = {2},
          eid = {50},
        pages = {50},
          doi = {10.3847/1538-4357/ad40aa},
archivePrefix = {arXiv},
       eprint = {2401.17378},
 primaryClass = {astro-ph.HE},
       adsurl = {https://ui.adsabs.harvard.edu/abs/2024ApJ...968...50P}
}

@ARTICLE{2021ApJ...923....1P,
       author = {{Pleunis}, Ziggy and {Good}, Deborah C. and {Kaspi}, Victoria M. and {Mckinven}, Ryan and {Ransom}, Scott M. and {Scholz}, Paul and {Bandura}, Kevin and {Bhardwaj}, Mohit and {Boyle}, P.~J. and {Brar}, Charanjot and {Cassanelli}, Tomas and {Chawla}, Pragya and {(Adam) Dong}, Fengqiu and {Fonseca}, Emmanuel and {Gaensler}, B.~M. and {Josephy}, Alexander and {Kaczmarek}, Jane F. and {Leung}, Calvin and {Lin}, Hsiu-Hsien and {Masui}, Kiyoshi W. and {Mena-Parra}, Juan and {Michilli}, Daniele and {Ng}, Cherry and {Patel}, Chitrang and {Rafiei-Ravandi}, Masoud and {Rahman}, Mubdi and {Sanghavi}, Pranav and {Shin}, Kaitlyn and {Smith}, Kendrick M. and {Stairs}, Ingrid H. and {Tendulkar}, Shriharsh P.},
        title = "{Fast Radio Burst Morphology in the First CHIME/FRB Catalog}",
      journal = {\apj},
         year = 2021,
        month = dec,
       volume = {923},
       number = {1},
          eid = {1},
        pages = {1},
          doi = {10.3847/1538-4357/ac33ac},
archivePrefix = {arXiv},
       eprint = {2106.04356},
 primaryClass = {astro-ph.HE},
       adsurl = {https://ui.adsabs.harvard.edu/abs/2021ApJ...923....1P}
}

@ARTICLE{2025ApJ...979..160S,
       author = {{Sand}, Ketan R. and {Curtin}, Alice P. and {Michilli}, Daniele and {Kaspi}, Victoria M. and {Fonseca}, Emmanuel and {Nimmo}, Kenzie and {Pleunis}, Ziggy and {Shin}, Kaitlyn and {Bhardwaj}, Mohit and {Brar}, Charanjot and {Dobbs}, Matt and {Eadie}, Gwendolyn M. and {Gaensler}, B.~M. and {Joseph}, Ronniy C. and {Leung}, Calvin and {Main}, Robert and {Masui}, Kiyoshi W. and {Mckinven}, Ryan and {Pandhi}, Ayush and {Pearlman}, Aaron B. and {Rafiei-Ravandi}, Masoud and {Sammons}, Mawson W. and {Smith}, Kendrick and {Stairs}, Ingrid H.},
        title = "{Morphology of 137 Fast Radio Bursts Down to Microsecond Timescales from the First CHIME/FRB Baseband Catalog}",
      journal = {\apj},
         year = 2025,
        month = feb,
       volume = {979},
       number = {2},
          eid = {160},
        pages = {160},
          doi = {10.3847/1538-4357/ad9b11},
archivePrefix = {arXiv},
       eprint = {2408.13215},
 primaryClass = {astro-ph.HE},
       adsurl = {https://ui.adsabs.harvard.edu/abs/2025ApJ...979..160S}
}

@ARTICLE{1975ApJ...197..185R,
       author = {{Rickett}, B.~J.},
        title = "{Amplitude-modulated noise: an empirical model for the radio radiation received from pulsars.}",
      journal = {\apj},
         year = 1975,
        month = apr,
       volume = {197},
        pages = {185-191},
          doi = {10.1086/153501},
       adsurl = {https://ui.adsabs.harvard.edu/abs/1975ApJ...197..185R}
}

@ARTICLE{1976ApJ...210..780C,
       author = {{Cordes}, J.~M.},
        title = "{Pulsar radiation as polarized shot noise.}",
      journal = {\apj},
         year = 1976,
        month = dec,
       volume = {210},
        pages = {780-791},
          doi = {10.1086/154887},
       adsurl = {https://ui.adsabs.harvard.edu/abs/1976ApJ...210..780C}
}

@ARTICLE{2011MNRAS.418.1258O,
       author = {{Os{\l}owski}, S. and {van Straten}, W. and {Hobbs}, G.~B. and {Bailes}, M. and {Demorest}, P.},
        title = "{High signal-to-noise ratio observations and the ultimate limits of precision pulsar timing}",
      journal = {\mnras},
         year = 2011,
        month = dec,
       volume = {418},
       number = {2},
        pages = {1258-1271},
          doi = {10.1111/j.1365-2966.2011.19578.x},
archivePrefix = {arXiv},
       eprint = {1108.0812},
 primaryClass = {astro-ph.GA},
       adsurl = {https://ui.adsabs.harvard.edu/abs/2011MNRAS.418.1258O}
}

@ARTICLE{2006ApJ...645..551M,
       author = {{McKinnon}, Mark M.},
        title = "{Orientation Angles of a Pulsar's Polarization Vector}",
      journal = {\apj},
         year = 2006,
        month = jul,
       volume = {645},
       number = {1},
        pages = {551-555},
          doi = {10.1086/504314},
archivePrefix = {arXiv},
       eprint = {astro-ph/0603446},
 primaryClass = {astro-ph},
       adsurl = {https://ui.adsabs.harvard.edu/abs/2006ApJ...645..551M}
}

@ARTICLE{1998ApJ...505..921M,
       author = {{Melrose}, D.~B. and {Macquart}, J. -P.},
        title = "{Stochastic Faraday Rotation}",
      journal = {\apj},
         year = 1998,
        month = oct,
       volume = {505},
       number = {2},
        pages = {921-927},
          doi = {10.1086/306204},
archivePrefix = {arXiv},
       eprint = {astro-ph/9805002},
 primaryClass = {astro-ph},
       adsurl = {https://ui.adsabs.harvard.edu/abs/1998ApJ...505..921M}
}

@ARTICLE{2004PASA...21..302H,
       author = {{Hotan}, A.~W. and {van Straten}, W. and {Manchester}, R.~N.},
        title = "{PSRCHIVE and PSRFITS: An Open Approach to Radio Pulsar Data Storage and Analysis}",
      journal = {\pasa},
         year = 2004,
        month = jan,
       volume = {21},
       number = {3},
        pages = {302-309},
          doi = {10.1071/AS04022},
archivePrefix = {arXiv},
       eprint = {astro-ph/0404549},
 primaryClass = {astro-ph},
       adsurl = {https://ui.adsabs.harvard.edu/abs/2004PASA...21..302H}
}

@ARTICLE{2021JOSS....6.2757H,
       author = {{Hazboun}, Jeffrey and {Shapiro-Albert}, Brent and {Baker}, Paul and {Henkel}, Amelia and {Wagner}, Cassidy and {Hesse}, Jacob and {Brook}, Paul and {Lam}, Michael and {McLaughlin}, Maura and {Garver-Daniels}, Nathan},
        title = "{The Pulsar Signal Simulator: A Python package for simulating radio signal data from pulsars}",
      journal = {The Journal of Open Source Software},
         year = 2021,
        month = feb,
       volume = {6},
       number = {58},
          eid = {2757},
        pages = {2757},
          doi = {10.21105/joss.02757},
       adsurl = {https://ui.adsabs.harvard.edu/abs/2021JOSS....6.2757H}
}

@ARTICLE{2020MNRAS.497.3335D,
       author = {{Day}, Cherie K. and {Deller}, Adam T. and {Shannon}, Ryan M. and {Qiu(邱昊)}, Hao and {Bannister}, Keith W. and {Bhandari}, Shivani and {Ekers}, Ron and {Flynn}, Chris and {James}, Clancy W. and {Macquart}, Jean-Pierre and {Mahony}, Elizabeth K. and {Phillips}, Chris J. and {Xavier Prochaska}, J.},
        title = "{High time resolution and polarization properties of ASKAP-localized fast radio bursts}",
      journal = {\mnras},
         year = 2020,
        month = sep,
       volume = {497},
       number = {3},
        pages = {3335-3350},
          doi = {10.1093/mnras/staa2138},
archivePrefix = {arXiv},
       eprint = {2005.13162},
 primaryClass = {astro-ph.HE},
       adsurl = {https://ui.adsabs.harvard.edu/abs/2020MNRAS.497.3335D}
}

@ARTICLE{2001ApJ...553..341E,
       author = {{Everett}, J.~E. and {Weisberg}, J.~M.},
        title = "{Emission Beam Geometry of Selected Pulsars Derived from Average Pulse Polarization Data}",
      journal = {\apj},
         year = 2001,
        month = may,
       volume = {553},
       number = {1},
        pages = {341-357},
          doi = {10.1086/320652},
archivePrefix = {arXiv},
       eprint = {astro-ph/0009266},
 primaryClass = {astro-ph},
       adsurl = {https://ui.adsabs.harvard.edu/abs/2001ApJ...553..341E}
}

@ARTICLE{2003A&A...410..253L,
       author = {{Li}, X.~H. and {Han}, J.~L.},
        title = "{The effect of scattering on pulsar polarization angle}",
      journal = {\aap},
         year = 2003,
        month = oct,
       volume = {410},
        pages = {253-256},
          doi = {10.1051/0004-6361:20031190},
archivePrefix = {arXiv},
       eprint = {astro-ph/0308095},
 primaryClass = {astro-ph},
       adsurl = {https://ui.adsabs.harvard.edu/abs/2003A&A...410..253L}
}

@ARTICLE{2021NatAs...5..594N,
       author = {{Nimmo}, K. and {Hessels}, J.~W.~T. and {Keimpema}, A. and {Archibald}, A.~M. and {Cordes}, J.~M. and {Karuppusamy}, R. and {Kirsten}, F. and {Li}, D.~Z. and {Marcote}, B. and {Paragi}, Z.},
        title = "{Highly polarized microstructure from the repeating FRB 20180916B}",
      journal = {Nature Astronomy},
         year = 2021,
        month = jun,
       volume = {5},
        pages = {594-603},
          doi = {10.1038/s41550-021-01321-3},
archivePrefix = {arXiv},
       eprint = {2010.05800},
 primaryClass = {astro-ph.HE},
       adsurl = {https://ui.adsabs.harvard.edu/abs/2021NatAs...5..594N}
}

@ARTICLE{2023MNRAS.526.2039H,
       author = {{Hewitt}, Dant{\'e} M. and {Hessels}, Jason W.~T. and {Ould-Boukattine}, Omar S. and {Chawla}, Pragya and {Cognard}, Isma{\"e}l and {Gopinath}, Akshatha and {Guillemot}, Lucas and {Huppenkothen}, Daniela and {Nimmo}, Kenzie and {Snelders}, Mark P.},
        title = "{Dense forests of microshots in bursts from FRB 20220912A}",
      journal = {\mnras},
         year = 2023,
        month = dec,
       volume = {526},
       number = {2},
        pages = {2039-2057},
          doi = {10.1093/mnras/stad2847},
archivePrefix = {arXiv},
       eprint = {2308.12118},
 primaryClass = {astro-ph.HE},
       adsurl = {https://ui.adsabs.harvard.edu/abs/2023MNRAS.526.2039H}
}

@ARTICLE{2022Sci...375.1266F,
       author = {{Feng}, Yi and {Li}, Di and {Yang}, Yuan-Pei and {Zhang}, Yongkun and {Zhu}, Weiwei and {Zhang}, Bing and {Lu}, Wenbin and {Wang}, Pei and {Dai}, Shi and {Lynch}, Ryan S. and {Yao}, Jumei and {Jiang}, Jinchen and {Niu}, Jiarui and {Zhou}, Dejiang and {Xu}, Heng and {Miao}, Chenchen and {Niu}, Chenhui and {Meng}, Lingqi and {Qian}, Lei and {Tsai}, Chao-Wei and {Wang}, Bojun and {Xue}, Mengyao and {Yue}, Youling and {Yuan}, Mao and {Zhang}, Songbo and {Zhang}, Lei},
        title = "{Frequency-dependent polarization of repeating fast radio bursts{\textemdash}implications for their origin}",
      journal = {Science},
         year = 2022,
        month = mar,
       volume = {375},
       number = {6586},
        pages = {1266-1270},
          doi = {10.1126/science.abl7759},
archivePrefix = {arXiv},
       eprint = {2202.09601},
 primaryClass = {astro-ph.HE},
       adsurl = {https://ui.adsabs.harvard.edu/abs/2022Sci...375.1266F}
}

@ARTICLE{2025arXiv250319749U,
       author = {{Uttarkar}, Pavan A. and {Shannon}, Ryan M. and {Gourdji}, Kelly and {Deller}, Adam T. and {Dial}, Tyson and {Glowacki}, Marcin and {Bera}, Apurba and {Gordon}, Alexa C. and {Ryder}, Stuart D. and {Tejos}, Nicolas and {Bhandari}, Shivani and {Wang}, Yuanming},
        title = "{A depolarisation census of ASKAP fast radio bursts}",
      journal = {arXiv e-prints},
         year = 2025,
        month = mar,
          eid = {arXiv:2503.19749},
        pages = {arXiv:2503.19749},
          doi = {10.48550/arXiv.2503.19749},
archivePrefix = {arXiv},
       eprint = {2503.19749},
 primaryClass = {astro-ph.HE},
       adsurl = {https://ui.adsabs.harvard.edu/abs/2025arXiv250319749U}
}

@ARTICLE{2024ApJ...969L..29B,
       author = {{Bera}, Apurba and {James}, Clancy W. and {Deller}, Adam T. and {Bannister}, Keith W. and {Shannon}, Ryan M. and {Scott}, Danica R. and {Gourdji}, Kelly and {Marnoch}, Lachlan and {Glowacki}, Marcin and {Ekers}, Ronald D. and {Ryder}, Stuart D. and {Dial}, Tyson},
        title = "{The Curious Case of Twin Fast Radio Bursts: Evidence for Neutron Star Origin?}",
      journal = {\apjl},
         year = 2024,
        month = jul,
       volume = {969},
       number = {2},
          eid = {L29},
        pages = {L29},
          doi = {10.3847/2041-8213/ad5966},
archivePrefix = {arXiv},
       eprint = {2406.13704},
 primaryClass = {astro-ph.HE},
       adsurl = {https://ui.adsabs.harvard.edu/abs/2024ApJ...969L..29B}
}

@ARTICLE{2015ApJ...802..130H,
       author = {{Hankins}, T.~H. and {Jones}, G. and {Eilek}, J.~A.},
        title = "{The Crab Pulsar at Centimeter Wavelengths. I. Ensemble Characteristics}",
      journal = {\apj},
         year = 2015,
        month = apr,
       volume = {802},
       number = {2},
          eid = {130},
        pages = {130},
          doi = {10.1088/0004-637X/802/2/130},
archivePrefix = {arXiv},
       eprint = {1502.00677},
 primaryClass = {astro-ph.HE},
       adsurl = {https://ui.adsabs.harvard.edu/abs/2015ApJ...802..130H}
}

@ARTICLE{2010A&A...524A..60J,
       author = {{Jessner}, A. and {Popov}, M.~V. and {Kondratiev}, V.~I. and {Kovalev}, Y.~Y. and {Graham}, D. and {Zensus}, A. and {Soglasnov}, V.~A. and {Bilous}, A.~V. and {Moshkina}, O.~A.},
        title = "{Giant pulses with nanosecond time resolution detected from the Crab pulsar at 8.5 and 15.1 GHz}",
      journal = {\aap},
         year = 2010,
        month = dec,
       volume = {524},
          eid = {A60},
        pages = {A60},
          doi = {10.1051/0004-6361/201014806},
archivePrefix = {arXiv},
       eprint = {1008.3992},
 primaryClass = {astro-ph.HE},
       adsurl = {https://ui.adsabs.harvard.edu/abs/2010A&A...524A..60J}
}

@ARTICLE{2025ApJ...988..175L,
       author = {{Liu}, Xiaohui and {Xu}, Heng and {Niu}, Jiarui and {Zhang}, Yongkun and {Jiang}, Jinchen and {Zhou}, Dejiang and {Han}, Jinlin and {Zhu}, Weiwei and {Lee}, Kejia and {Li}, Di and {Wang}, Wei-Yang and {Zhang}, Bing and {Chen}, Xuelei and {Luo}, Jia-Wei and {Luo}, Rui and {Niu}, Chenhui and {Qu}, Yuanhong and {Wang}, Bojun and {Wang}, Fayin and {Wang}, Pei and {Wang}, Tiancong and {Wu}, Qin and {Wu}, Ziwei and {Xu}, Jiangwei and {Yang}, Yuan-Pei and {Zhang}, Jun-Shuo},
        title = "{Polarization Position Angle Swing and the Rotating Vector Model of Repeating Fast Radio Bursts}",
      journal = {\apj},
         year = 2025,
        month = aug,
       volume = {988},
       number = {2},
          eid = {175},
        pages = {175},
          doi = {10.3847/1538-4357/ade689},
archivePrefix = {arXiv},
       eprint = {2504.00391},
 primaryClass = {astro-ph.HE},
       adsurl = {https://ui.adsabs.harvard.edu/abs/2025ApJ...988..175L}
}

@ARTICLE{2009ApJ...694.1413V,
       author = {{van Straten}, W.},
        title = "{The Statistics of Radio Astronomical Polarimetry: Bright Sources and High Time Resolution}",
      journal = {\apj},
         year = 2009,
        month = apr,
       volume = {694},
       number = {2},
        pages = {1413-1422},
          doi = {10.1088/0004-637X/694/2/1413},
archivePrefix = {arXiv},
       eprint = {0812.3461},
 primaryClass = {astro-ph},
       adsurl = {https://ui.adsabs.harvard.edu/abs/2009ApJ...694.1413V}
}

@ARTICLE{2019MNRAS.490L..12B,
       author = {{Bera}, Apurba and {Chengalur}, Jayaram N.},
        title = "{Super-giant pulses from the Crab pulsar: energy distribution and occurrence rate}",
      journal = {\mnras},
         year = 2019,
        month = nov,
       volume = {490},
       number = {1},
        pages = {L12-L16},
          doi = {10.1093/mnrasl/slz140},
archivePrefix = {arXiv},
       eprint = {1909.13812},
 primaryClass = {astro-ph.HE},
       adsurl = {https://ui.adsabs.harvard.edu/abs/2019MNRAS.490L..12B}
}

@ARTICLE{2014MNRAS.442L...9L,
       author = {{Lyubarsky}, Yu.},
        title = "{A model for fast extragalactic radio bursts.}",
      journal = {\mnras},
         year = 2014,
        month = jul,
       volume = {442},
        pages = {L9-L13},
          doi = {10.1093/mnrasl/slu046},
archivePrefix = {arXiv},
       eprint = {1401.6674},
 primaryClass = {astro-ph.HE},
       adsurl = {https://ui.adsabs.harvard.edu/abs/2014MNRAS.442L...9L}
}

@ARTICLE{2018MNRAS.477.2470L,
       author = {{Lu}, Wenbin and {Kumar}, Pawan},
        title = "{On the radiation mechanism of repeating fast radio bursts}",
      journal = {\mnras},
         year = 2018,
        month = jun,
       volume = {477},
       number = {2},
        pages = {2470-2493},
          doi = {10.1093/mnras/sty716},
archivePrefix = {arXiv},
       eprint = {1710.10270},
 primaryClass = {astro-ph.HE},
       adsurl = {https://ui.adsabs.harvard.edu/abs/2018MNRAS.477.2470L}
}

@ARTICLE{2022MNRAS.510.4654B,
       author = {{Beniamini}, Paz and {Kumar}, Pawan and {Narayan}, Ramesh},
        title = "{Faraday depolarization and induced circular polarization by multipath propagation with application to FRBs}",
      journal = {\mnras},
         year = 2022,
        month = mar,
       volume = {510},
       number = {3},
        pages = {4654-4668},
          doi = {10.1093/mnras/stab3730},
archivePrefix = {arXiv},
       eprint = {2110.00028},
 primaryClass = {astro-ph.HE},
       adsurl = {https://ui.adsabs.harvard.edu/abs/2022MNRAS.510.4654B}
}

@ARTICLE{2026MNRAS.545f1997U,
       author = {{Uttarkar}, Pavan A. and {Shannon}, Ryan M. and {Gourdji}, Kelly and {Deller}, Adam T. and {Dial}, Tyson and {Glowacki}, Marcin and {Bera}, Apurba and {Gordon}, Alexa C. and {Ryder}, Stuart D. and {Tejos}, Nicolas and {Bhandari}, Shivani and {Wang}, Yuanming},
        title = "{A depolarization census of ASKAP fast radio bursts}",
      journal = {\mnras},
         year = 2026,
        month = jan,
       volume = {545},
       number = {2},
          eid = {staf1997},
        pages = {staf1997},
          doi = {10.1093/mnras/staf1997},
archivePrefix = {arXiv},
       eprint = {2503.19749},
 primaryClass = {astro-ph.HE},
       adsurl = {https://ui.adsabs.harvard.edu/abs/2026MNRAS.545f1997U}
}

@ARTICLE{2025PASA...42...36S,
       author = {{Shannon}, Ryan M. and {Bannister}, Keith W. and {Bera}, Apurba and {Bhandari}, Shivani and {Day}, Cherie K. and {Deller}, Adam T. and {Dial}, Tyson and {Dobie}, Dougal and {Ekers}, Ron D. and {Fong}, Wen-fai and {Glowacki}, Marcin and {Gordon}, Alexa C. and {Gourdji}, Kelly and {Jaini}, Akhil and {James}, Clancy W. and {Kumar}, Pravir and {Mahony}, Elizabeth K. and {Marnoch}, Lachlan and {Muller}, August R. and {Prochaska}, Xavier and {Qiu}, Hao and {Ryder}, Stuart D. and {Sadler}, Elaine M. and {Scott}, Danica R. and {Tejos}, N. and {Uttarkar}, Pavan A. and {Wang}, Yuanming},
        title = "{The commensal real-time ASKAP fast transient incoherent-sum survey}",
      journal = {\pasa},
         year = 2025,
        month = jan,
       volume = {42},
          eid = {e036},
        pages = {e036},
          doi = {10.1017/pasa.2025.8},
archivePrefix = {arXiv},
       eprint = {2408.02083},
 primaryClass = {astro-ph.HE},
       adsurl = {https://ui.adsabs.harvard.edu/abs/2025PASA...42...36S}
}

\appendix

\section{Back-Of-The-Envelope Derivation of the Expected Variance of the Polarisation Angle}\label{A1}
We give a sanity check for the expected variance of the polarisation angle (PA) under the microshot model introduced in the main text. Let the envelope have FWHM $W_0$ (ms) and each microshot have FWHM $w_i$, with fractional width $w_i/W_0 \in [w_{\min}, w_{\max}]$. For an order-of-magnitude estimate we replace the fractional width by its mean
\begin{equation*}
    \bar{w} \;=\; \tfrac{1}{2}\!\left(w_{\min}+w_{\max}\right), 
  \qquad
    w \equiv \bar{w}\,W_0.
\end{equation*}
Write $\kappa \equiv 2\sqrt{2\log{2}}$ so that FWHM$\to\sigma$ via $\sigma = \text{FWHM}/\kappa$. A thin-screen exponential scattering response with timescale $\tau_0$ (ms) broadens Gaussian widths in quadrature,
\begin{equation*}
  \sigma_{W,\mathrm{tot}} \;=\; \sqrt{\sigma_W^2 + \tau_0^2},
  \qquad
  \sigma_{w,\mathrm{tot}} \;=\; \sqrt{\sigma_w^2 + \tau_0^2},
\end{equation*}
with $\sigma_W = W_0/\kappa$ and $\sigma_w = w/\kappa$. The corresponding broadened FWHM measures are
\begin{equation*}
  W_{\mathrm{tot}} \;=\; \kappa\,\sigma_{W,\mathrm{tot}},
  \qquad
  w_{\mathrm{tot}} \;=\; \kappa\,\sigma_{w,\mathrm{tot}}.
\end{equation*}

If $N$ microshots occur within the envelope, the number that effectively contribute at any time scales with the broadened-width ratio. Defining
\begin{equation*}
  r \;\equiv\; \frac{w_{\mathrm{tot}}}{W_{\mathrm{tot}}}
  \;=\; \frac{\sigma_{w,\mathrm{tot}}}{\sigma_{W,\mathrm{tot}}}, 
  \qquad
  N_{\mathrm{eff}} \;=\; N\,r,
\end{equation*}
and letting $\sigma_\psi$ be the intrinsic single-shot PA scatter (radians), the rms PA scatter is
\begin{equation}\label{eq:psirms_basic}
  \psi_{\mathrm{rms}} \;\approx\; \frac{\sigma_\psi}{\sqrt{N_{\mathrm{eff}}}},
\end{equation}
so the basic variance estimate is
\begin{equation}\label{eq:var_basic}
  \mathbb{V}_{\exp}(\psi)
  \;=\; \psi_{\mathrm{rms}}^2
  \;=\; \frac{\sigma_\psi^2}{N\,r}
  \;=\; \frac{\sigma_\psi^2\,W_{\mathrm{tot}}}{N\,w_{\mathrm{tot}}}.
\end{equation}

When estimating fluctuations about the mean PA across the envelope, subtracting the mean introduces a degrees-of-freedom correction. Approximating the number of independent samples by $M \approx W_{\mathrm{tot}}/w_{\mathrm{tot}} = 1/r$ gives $(M-1)/M \approx 1-r$, yielding
\begin{equation}\label{eq:var_basic_corr}
  \mathbb{V}_{\exp}(\psi)
  \;=\; \frac{\sigma_\psi^2}{N\,r}\,\bigl(1-r\bigr)
  \;=\; \frac{\sigma_\psi^2\,W_{\mathrm{tot}}}{N\,w_{\mathrm{tot}}}\!
  \left(1 - \frac{w_{\mathrm{tot}}}{W_{\mathrm{tot}}}\right).
\end{equation}
Simplifying and converting to deg.$^2$,
\begin{equation}
    \mathbb{V}_{\exp}(\psi) \;=\; 
    \frac{\sigma_\psi^2}{N}\!
    \left(\frac{W_{\mathrm{tot}}}{w_{\mathrm{tot}}} - 1\right)\left(\frac{180}{\pi}\right)^2.
\end{equation}

This corrected estimate tends to zero as scattering grows large ($r\to 1$). It ignores amplitude weighting, detailed temporal structure, and the full width distribution, but provides a rapid order-of-magnitude check. If needed for discretely sampled data with time resolution $\delta t$, one may downweight $N_{\mathrm{eff}}$ by $\min\!\bigl(1,\,\sigma_{w,\mathrm{tot}}/(\delta t/\kappa)\bigr)$ to account for unresolved microshots.

\section{FRB~20240318A FIRES Recreation and Data}
\begin{figure}[H]
    \centering
    \begin{subfigure}[b]{0.9\textwidth}
        \includegraphics[width=\textwidth]{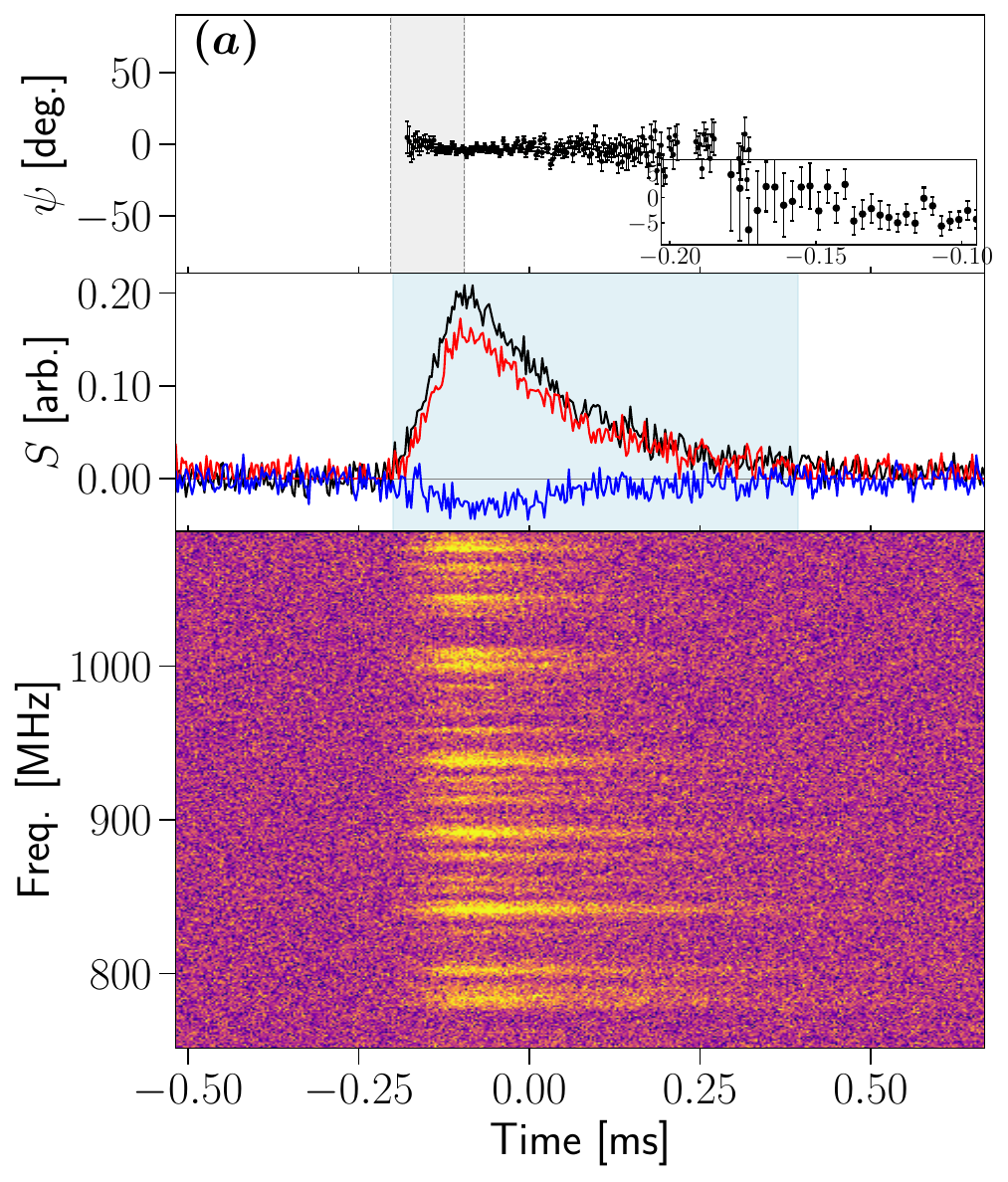}
        \captionlistentry{}\label{fig:240318A}
    \end{subfigure}\\
    \begin{subfigure}[b]{0.9\textwidth}
        \includegraphics[width=\textwidth]{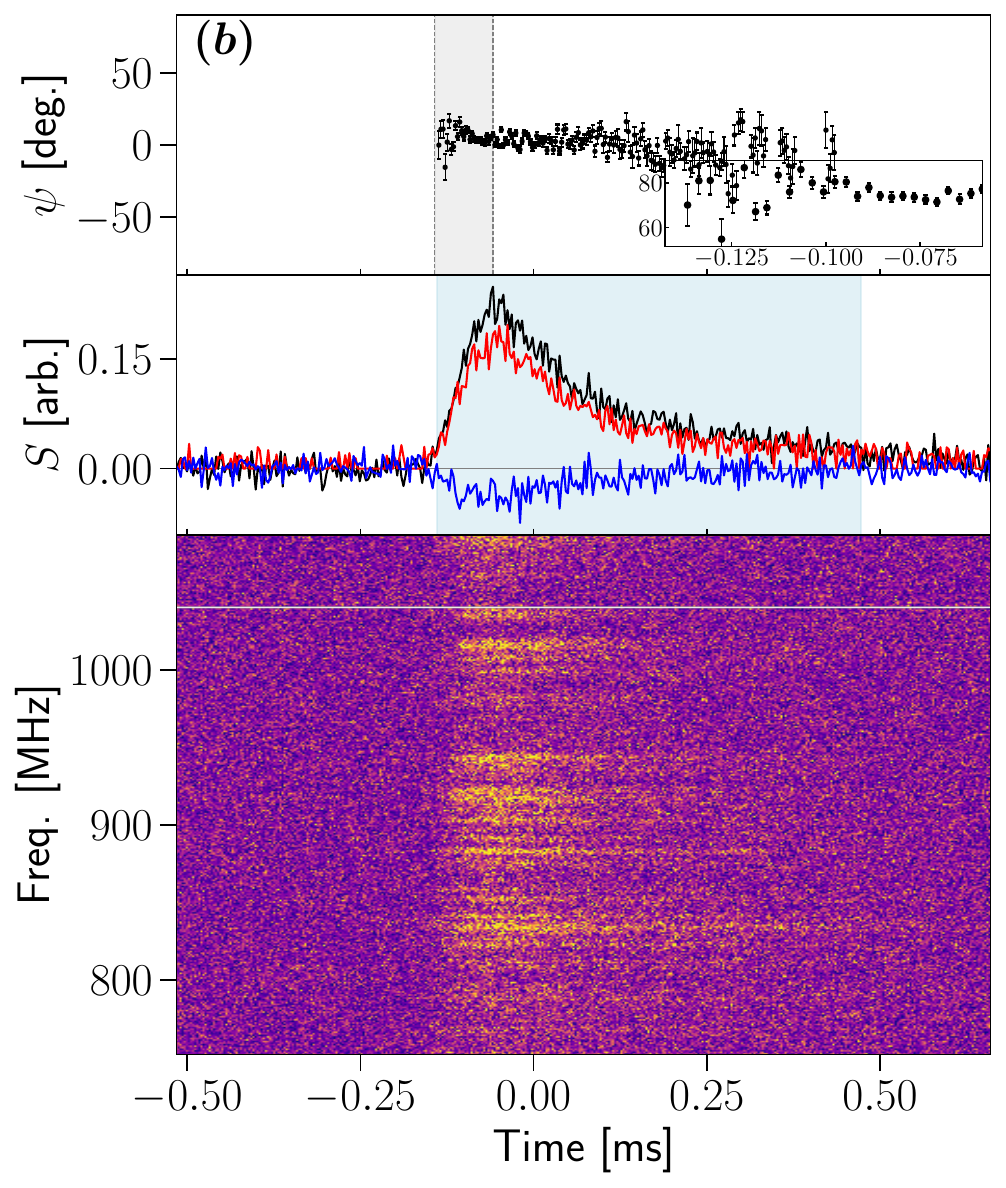}
        \captionlistentry{}\label{fig:HTR 240318A}
    \end{subfigure}
    \caption{A \texttt{FIRES} comparison with FRB~20240318A. Top: A \texttt{FIRES} recreation of FRB~20240318A comprised of 100 microshots, $\tau_{1\,\rm GHz} = 0.128$\,ms and scintillation (on-pulse S/N$=109$). Bottom: real FRB~20240318A data reproduced from~\cite{2025arXiv250517497S} (on-pulse S/N$\sim110$, RM corrected from RM=-48.03\,rad\,m$^{-2}$). The full list of parameters used are presented in Table~\ref{tab:params} and are described in Section~\ref{sec:model}. The top panels show the polarisation angle, the centre panels show the pulse profile, and the bottom panels show the dynamic spectra. The blue shaded regions in the centre panels is the minimum boxcar width that contains 95\% of the total flux in the pulse profile.}
    \label{fig:FRB_240318A_compare}
\end{figure}


\section{FRB 20240318A Parameter Distribution Comparisons}
\subsection{Amplitude Distributions}\label{subsec: amp dist}

\begin{figure*}[!p]
    \centering
    \begin{subfigure}[t]{0.45\textwidth}
        \centering
        \includegraphics[width=\linewidth]{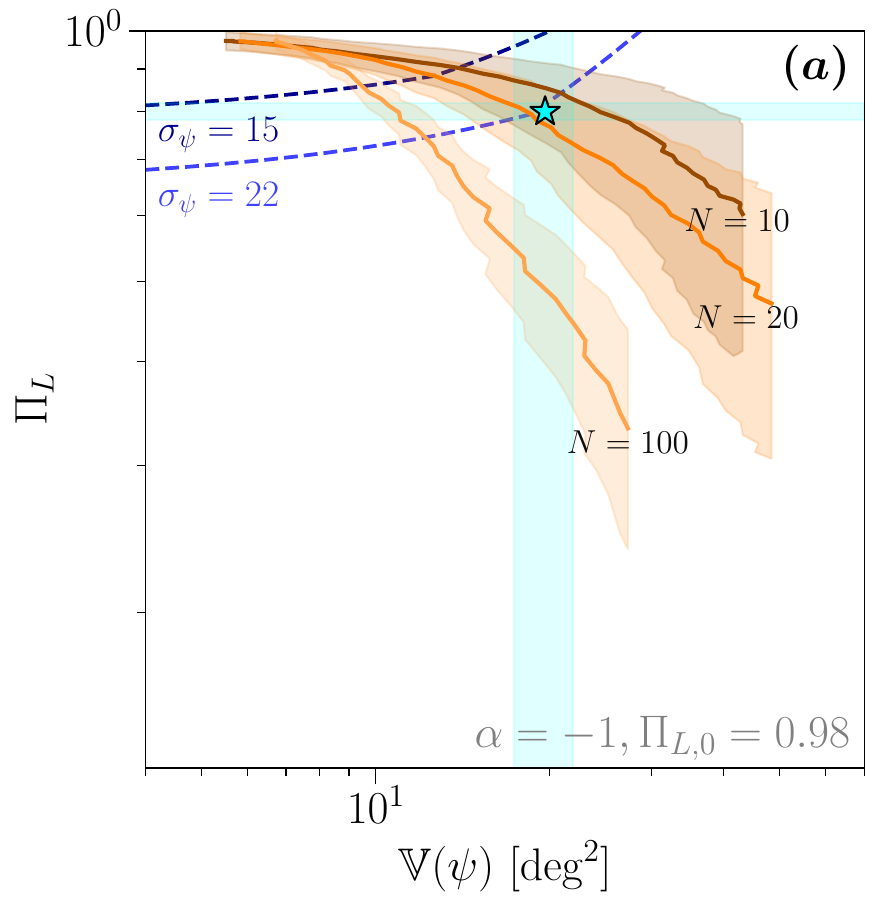}
        \captionlistentry{}\label{fig:lfrac_amps_P1_L0.98}
    \end{subfigure}
    \begin{subfigure}[t]{0.45\textwidth}
        \centering
        \includegraphics[width=\linewidth]{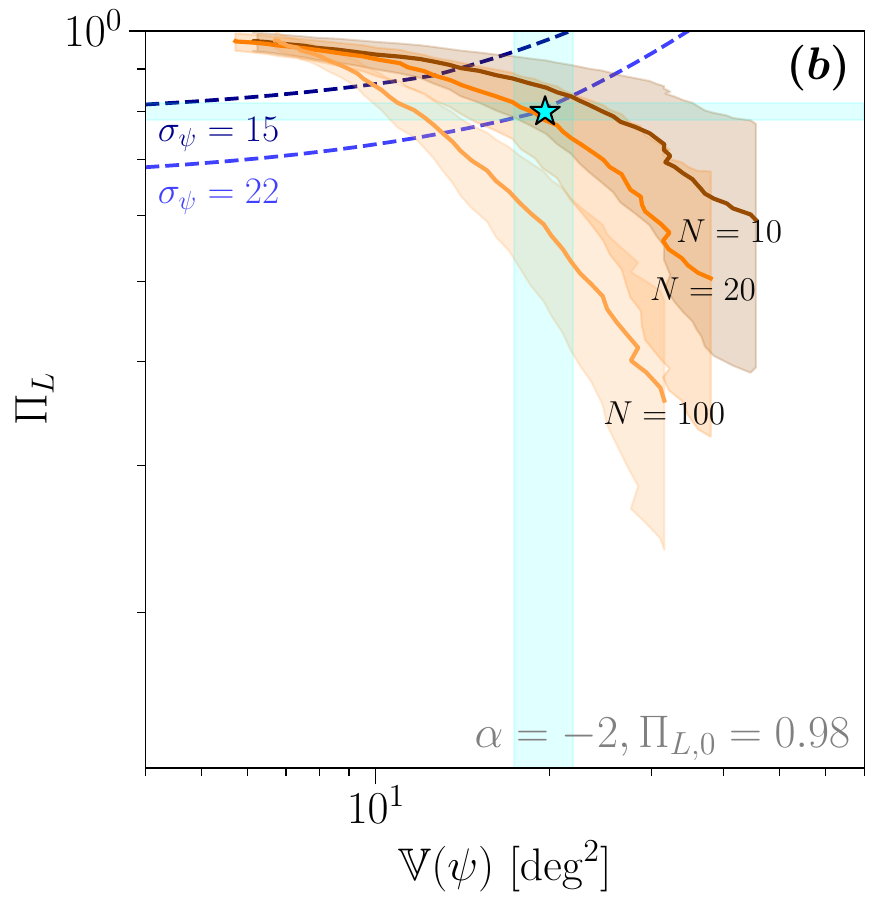}
        \captionlistentry{}\label{fig:lfrac_amps_P2_L0.98}
    \end{subfigure}\vspace{-43pt}\\

    \begin{subfigure}[t]{0.45\textwidth}
        \centering
        \includegraphics[width=\linewidth]{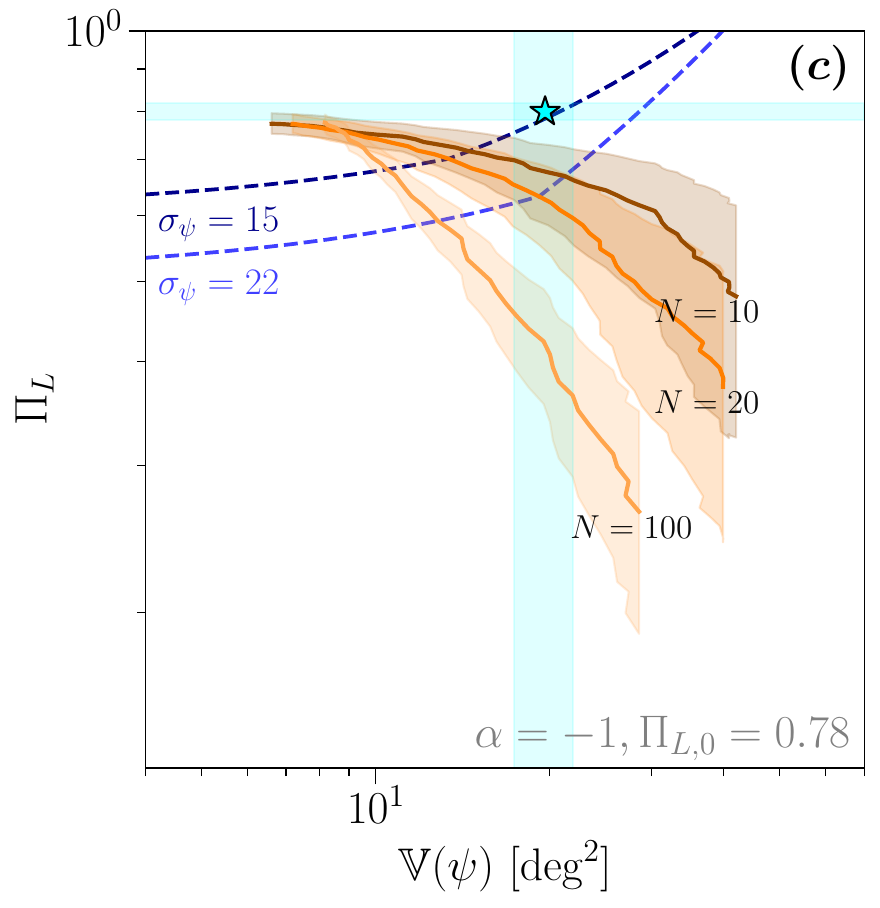}
        \captionlistentry{}\label{fig:lfrac_amps_P1_L0.78}
    \end{subfigure}
    \begin{subfigure}[t]{0.45\textwidth}
        \centering
        \includegraphics[width=\linewidth]{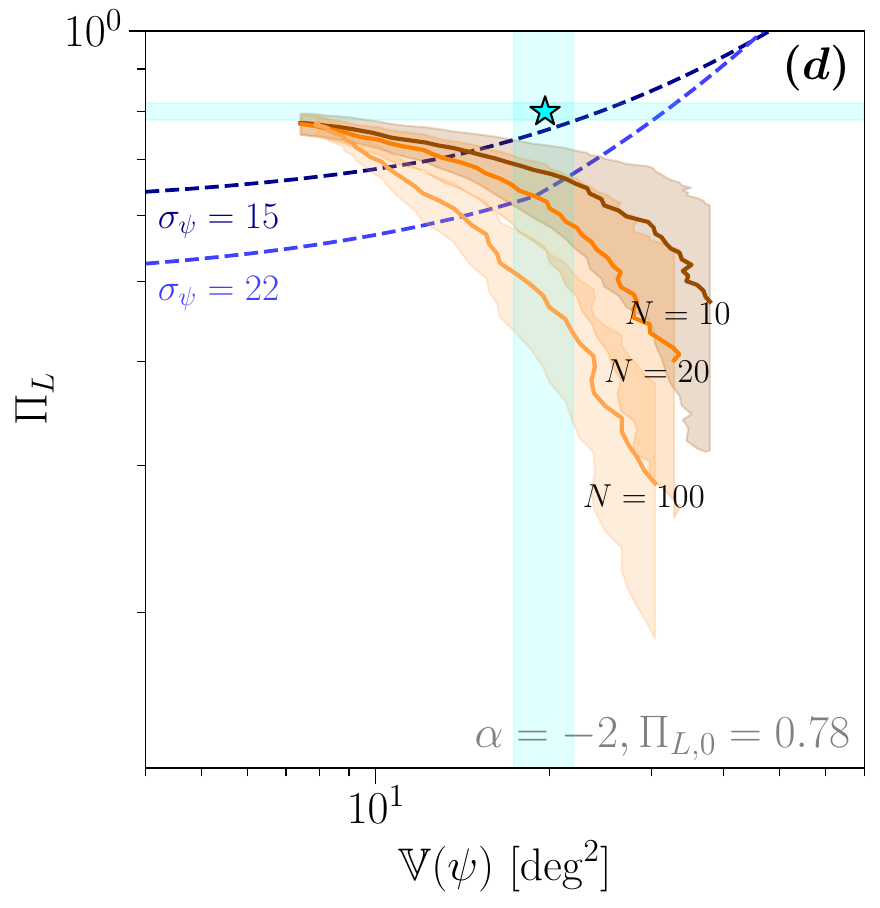}
        \captionlistentry{}\label{fig:lfrac_amps_P2_L0.78}
    \end{subfigure}

    \caption{Measured linear polarisation fraction $\Pi_{L}$ versus PA variance $\mathbb{V}(\psi)$ for mock FRB~20240318A power law amplitude distribution comparison. Left: $\alpha=-1$. Right: $\alpha=-2$. Top/bottom rows: intrinsic linear polarisation fraction $\Pi_{L,0}=0.98$ and $0.78$. Lines show medians for $N=10,20,100$ (500 realisations per $\sigma_\psi$); shaded regions are the 16th--84th percentiles. Cyan star: measured FRB~20240318A (S/N$\sim110$) with off-pulse RMS uncertainty. Blue dashed lines: loci of constant $\sigma_\psi$.}
    \label{fig:lfrac_amps_240318A_combined}
\end{figure*}

Here we compare power law amplitude distributions with $\alpha=-1 \textrm{ and } -2$. Figure~\ref{fig:lfrac_amps_240318A_combined} shows that the distributions aren't strongly distinguishable in the $\Pi_L$--$\mathbb{V}(\psi)$ plane, with the $\alpha=-1$ distribution producing slightly lower $\Pi_L$ and $\mathbb{V}(\psi)$ at fixed $\sigma_\psi$ and $N > 20$. This is because the $\alpha=-1$ distribution has more high-amplitude microshots that can dominate the polarisation properties. The differences between these distributions and the $\alpha=-3$ distribution used in Section~\ref{subsec:linpol} are relatively subtle compared to the effects of varying $N$ and $\sigma_\psi$, and all three distributions can reproduce the observed properties of FRB~20240318A within the uncertainties. We thus adopt the $\alpha=-3$ distribution for our main analysis as it is more consistent with observed Crab GP energy distributions~\citep{2019MNRAS.490L..12B}.

\subsection{Width Distributions}\label{subsec: width dist}

\begin{figure*}[!htbp]
    \centering

    \begin{subfigure}[t]{0.45\textwidth}
        \centering
        \includegraphics[width=\linewidth]{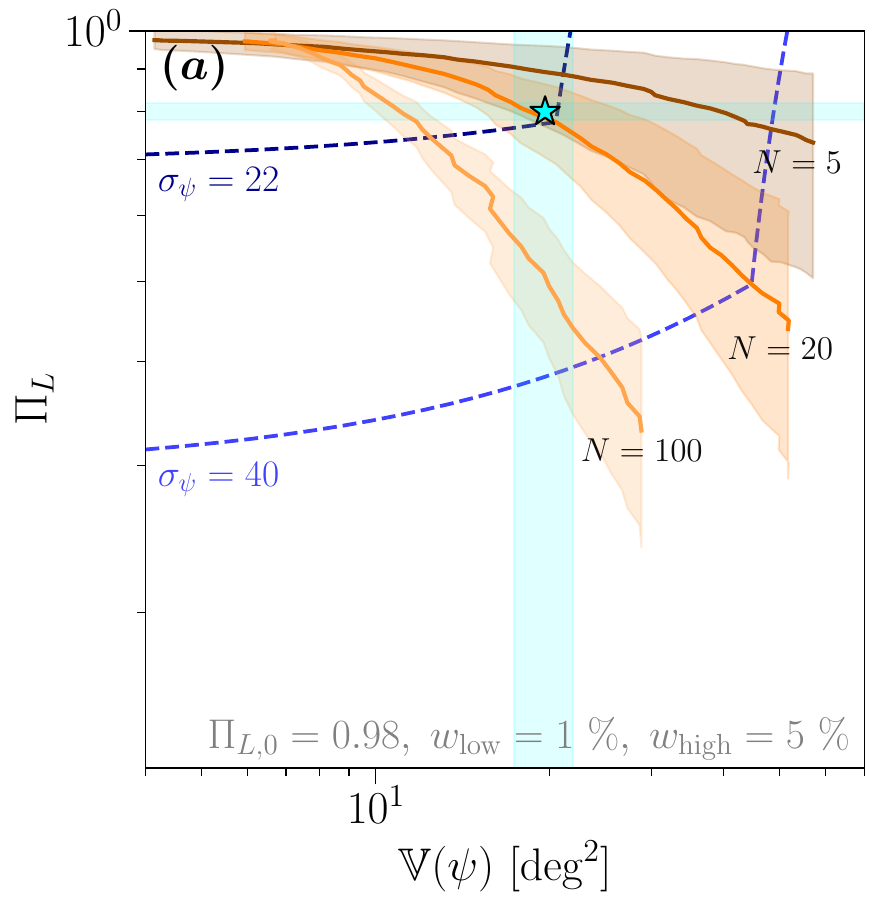}\par\medskip
        \captionlistentry{}\label{fig:lfrac_w1-5_240318A_L0.98}
    \end{subfigure}
    \begin{subfigure}[t]{0.45\textwidth}
        \centering
        \includegraphics[width=\linewidth]{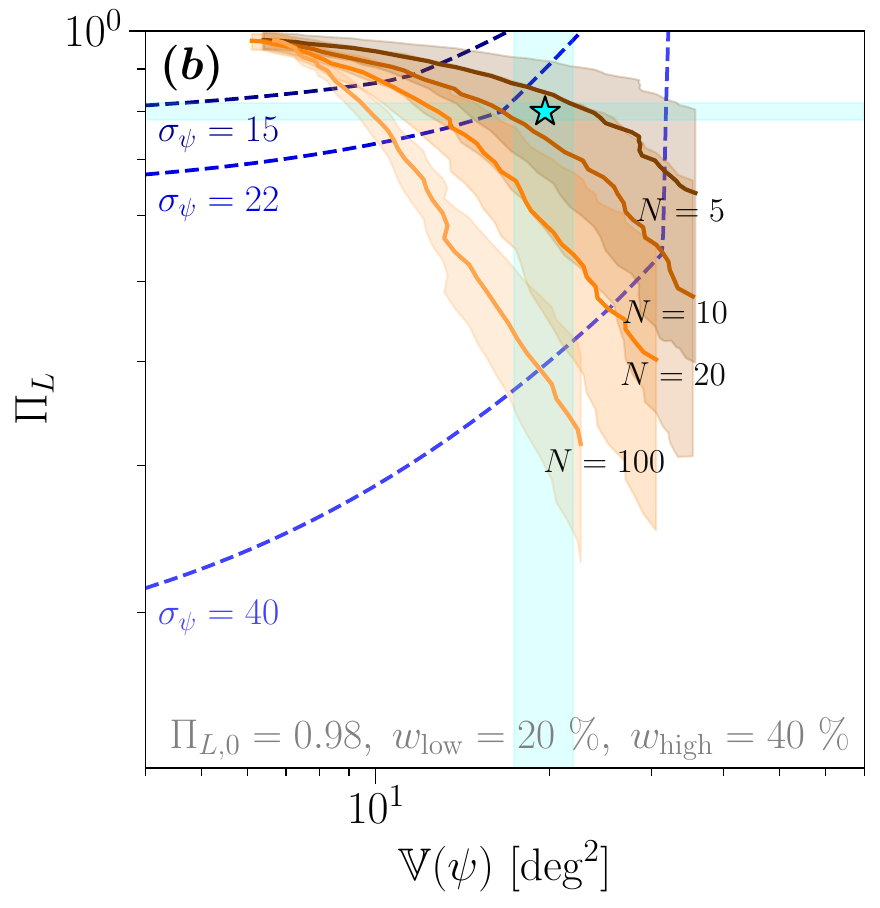}\par\medskip
        \captionlistentry{}\label{fig:lfrac_w20-40_240318A_L0.98}
    \end{subfigure}\vspace{-40pt}\\

    \begin{subfigure}[t]{0.45\textwidth}
        \centering
        \includegraphics[width=\linewidth]{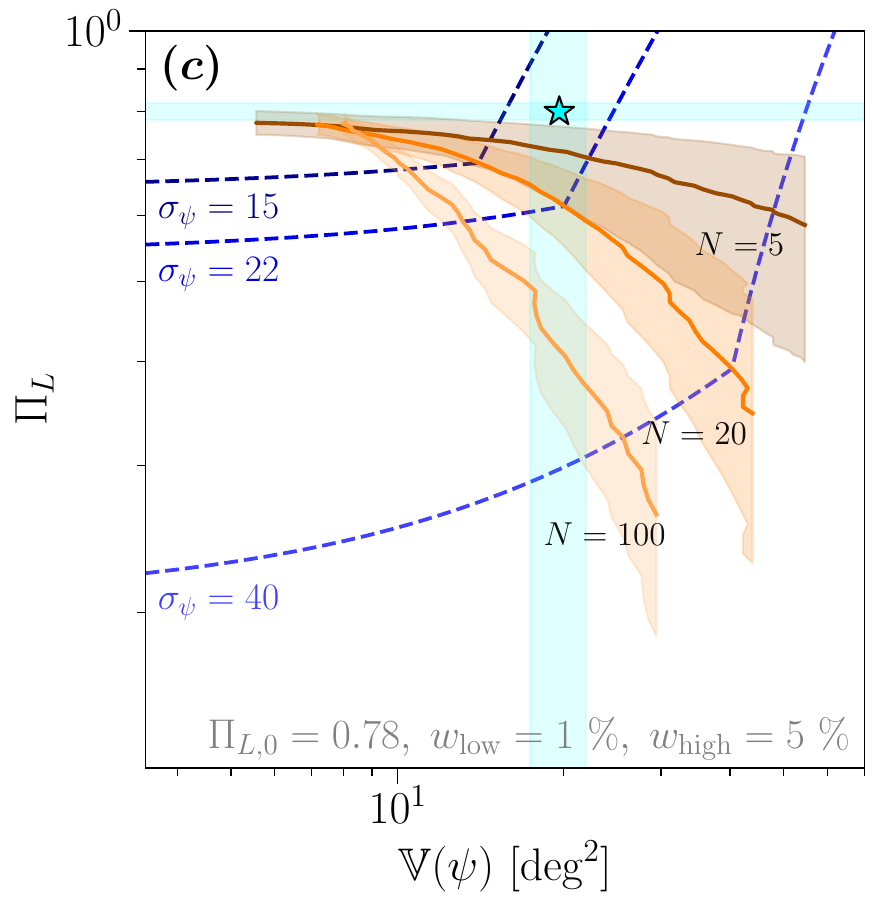}\par\medskip
        \captionlistentry{}\label{fig:lfrac_w1-5_240318A_L0.78}
    \end{subfigure}
    \begin{subfigure}[t]{0.45\textwidth}
        \centering
        \includegraphics[width=\linewidth]{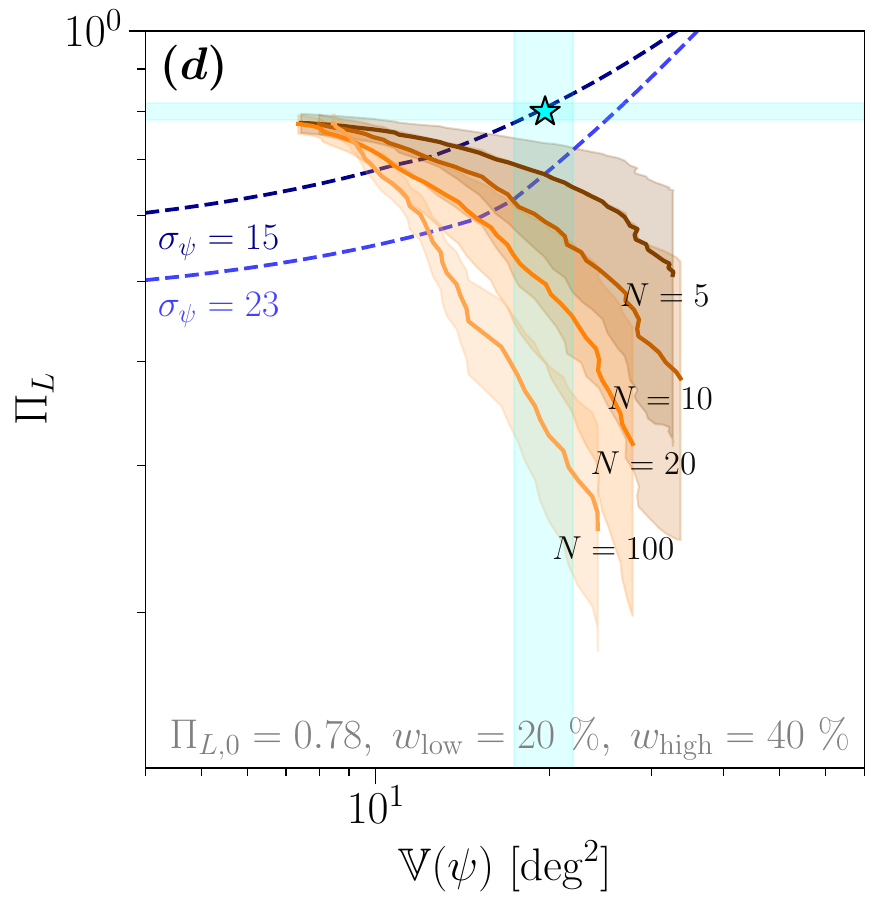}\par\medskip
        \captionlistentry{}\label{fig:lfrac_w20-40_240318A_L0.78}
    \end{subfigure}
    \caption{Comparison of width distributions for mock FRB~20240318A. Measured linear polarisation fraction $\Pi_{L}$ versus PA variance $\mathbb{V}(\psi)$. Left/right columns: microshot fractional FWHM in $[1\%,5\%]$ and $[20\%,40\%]$. Top/bottom rows: intrinsic microshot $\Pi_{L,0}=0.98$ and $0.78$. Lines show medians for $N=5,10,20,100$ (500 realisations per $\sigma_\psi$); shaded regions are the 16th--84th percentiles. Cyan star: measured FRB~20240318A (S/N$\sim110$) with off-pulse RMS uncertainty. Blue dashed lines: loci of constant $\sigma_\psi$. Some lines have been omitted for visual clarity.}\label{fig:lfrac_widths_240318A}
\end{figure*}

We note that varying microshot widths introduces a degeneracy to varying shot amplitudes. Here we compare simulations with width distributions of 1--5\% and 20--40\% of the burst width. Figure~\ref{fig:lfrac_widths_240318A} shows that the narrower width distribution produces a more extended track in the $\Pi_L$--$\mathbb{V}(\psi)$ plane, with the $N=5$ track extending to higher $\sigma_\psi$ values. The wider width distribution produces more compact tracks. This is because wider microshots are more likely to overlap and therefore produce more depolarisation and PA variance damping at fixed $N$ and $\sigma_\psi$. This degeneracy makes it difficult to distinguish between a scenario with fewer, wider microshots and one with more, narrower microshots.

\label{sec:FRB 20240318A widths}

\section{Correlations Between $\Delta\psi$--$\Delta\Pi_L$}\label{sec: PA L/I}
Our simulations for $\Delta\psi \equiv \psi-\bar{\psi}$ versus $\Delta\Pi_{L} \equiv \Pi_{L}-\bar{\Pi}_{L}$ (Figure~\ref{fig:pa-li-full}) show that depolarisation is governed by effective averaging (microshot overlap plus propagation/instrumental smoothing), rather than by intrinsic PA scatter alone.

\begin{figure*}[!htbp]
    \centering

    \begin{subfigure}[t]{0.45\textwidth}
        \centering
        \includegraphics[width=\linewidth]{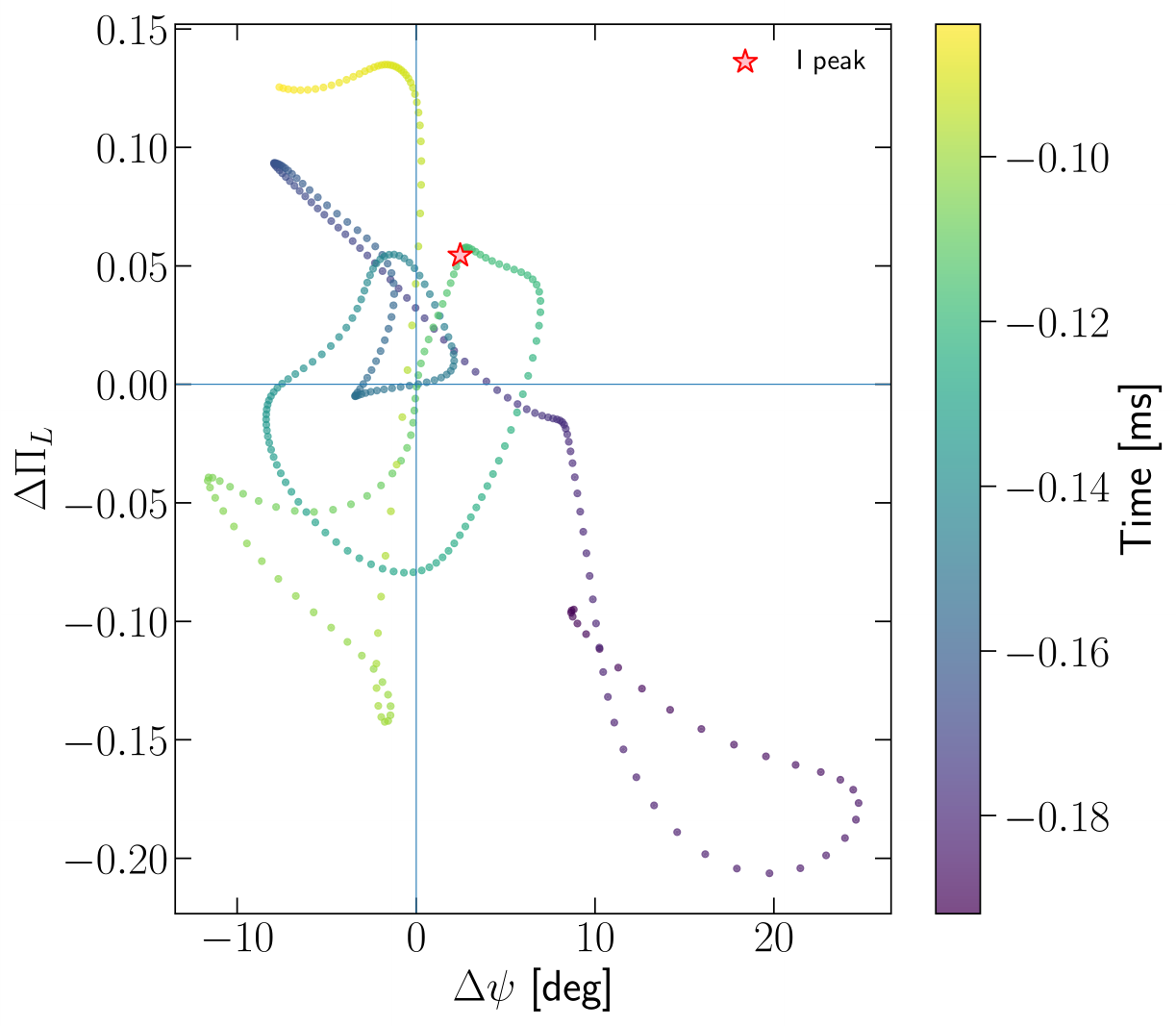}\par\medskip
        \captionlistentry{}\label{fig:pa-li-nn-ns}
    \end{subfigure}
    \begin{subfigure}[t]{0.45\textwidth}
        \centering
        \includegraphics[width=\linewidth, trim={0 0 0 1.5cm},clip]{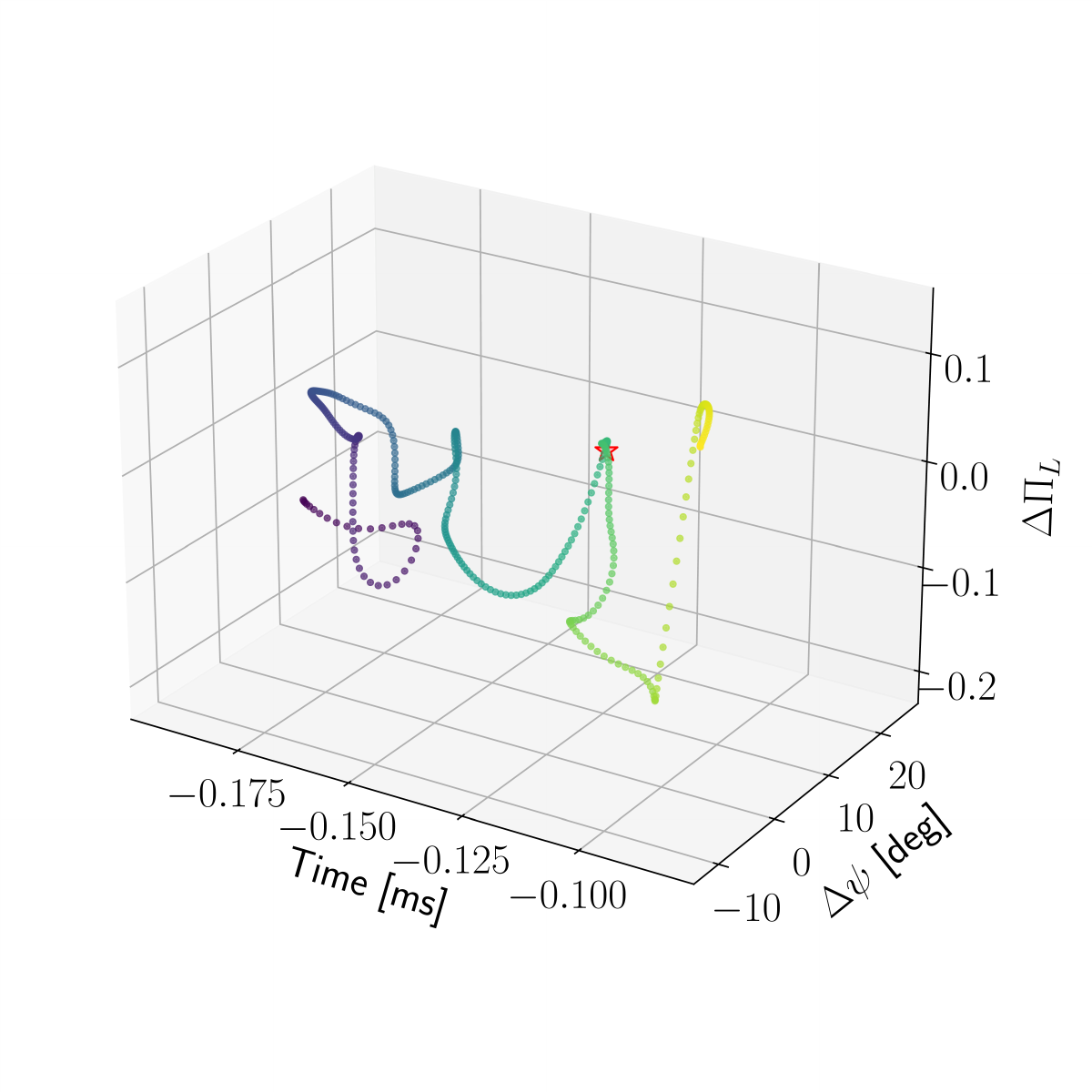}\par\medskip
        \captionlistentry{}\label{fig:pa-li-nn-ns-3d}
    \end{subfigure}\vspace{-20pt}\\

    \begin{subfigure}[t]{0.45\textwidth}
        \centering
        \includegraphics[width=\linewidth]{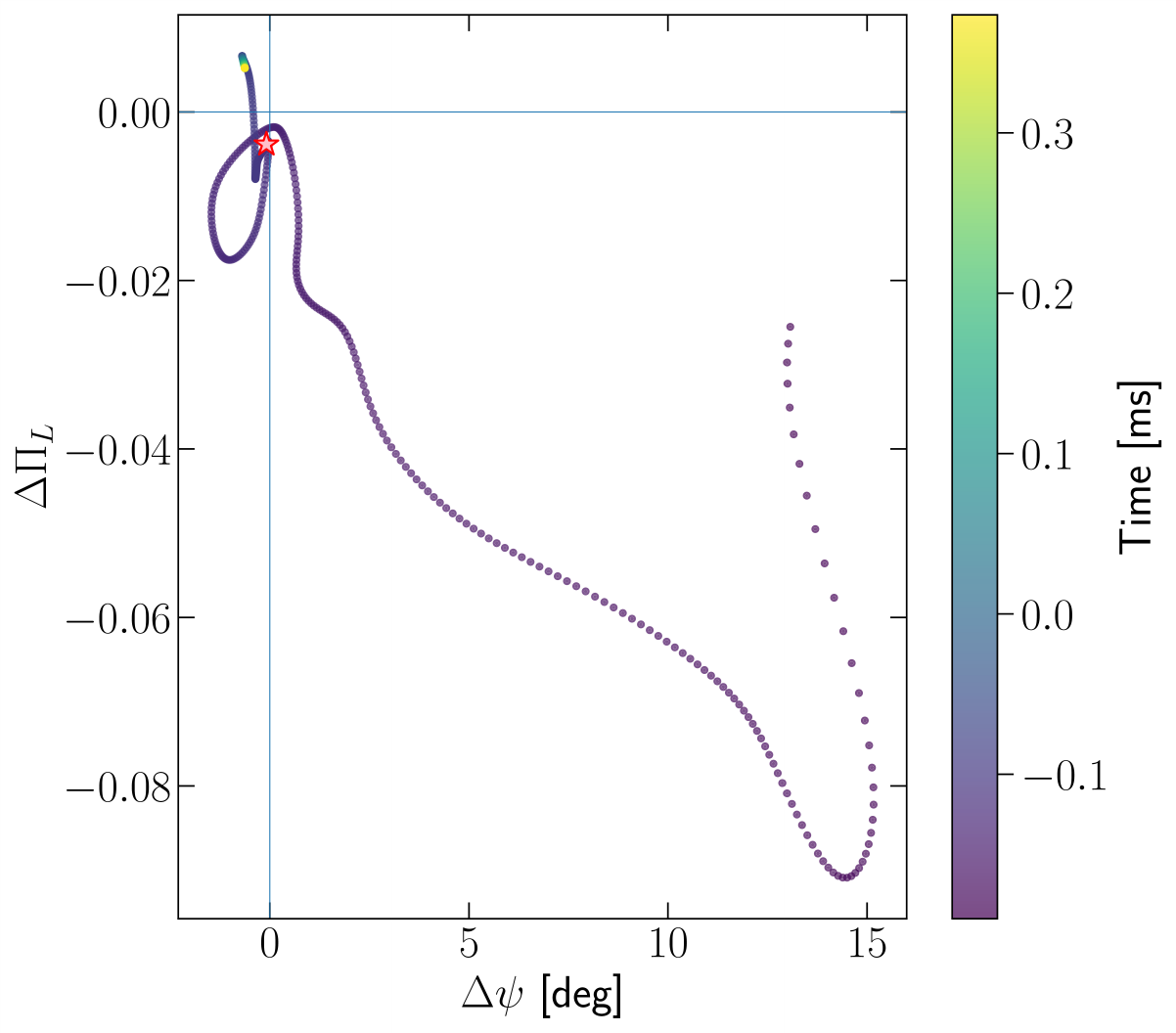}\par\medskip
        \captionlistentry{}\label{fig:pa-li-nn}
    \end{subfigure}
    \begin{subfigure}[t]{0.45\textwidth}
        \centering
        \includegraphics[width=\linewidth, trim={0 0 0 1.5cm},clip]{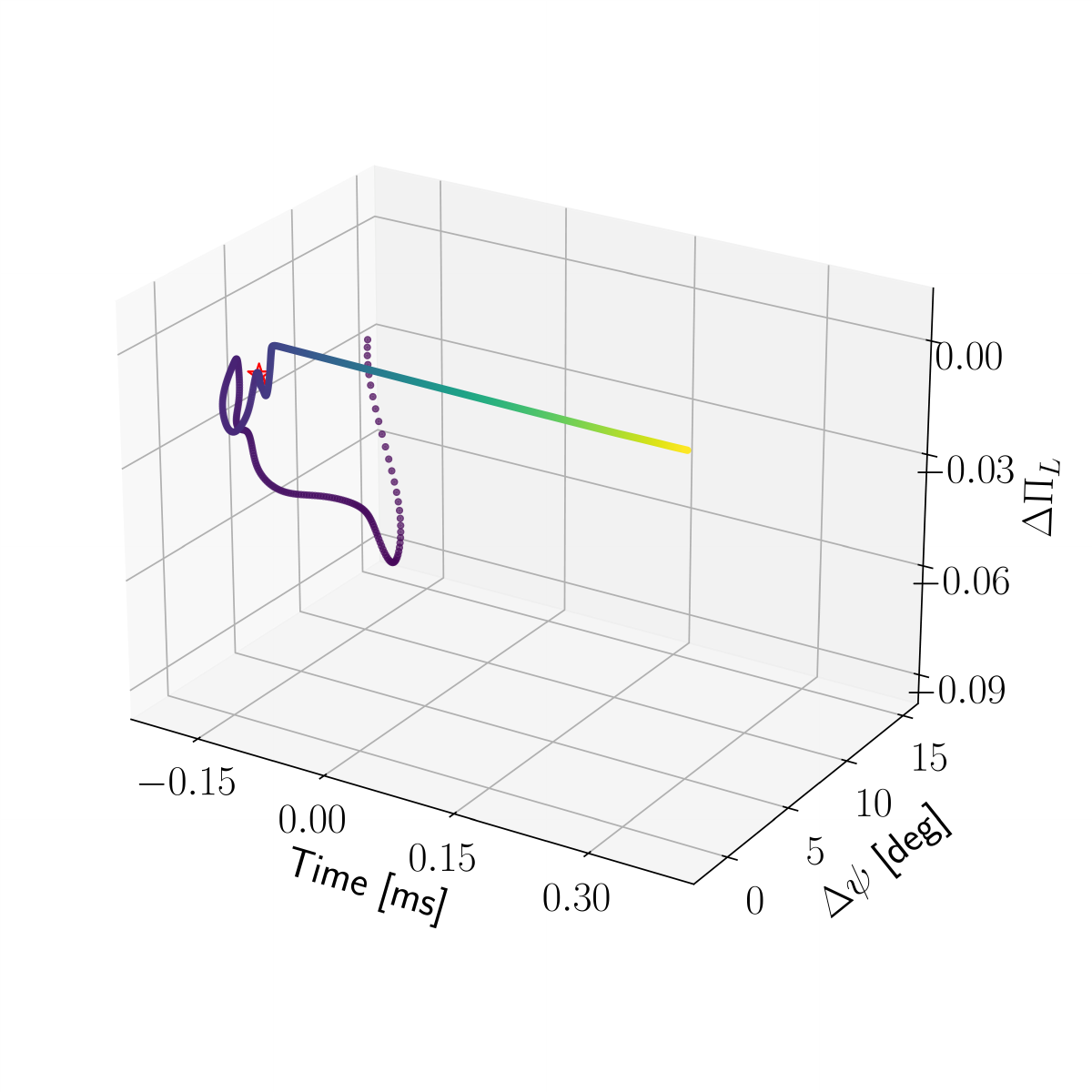}\par\medskip
        \captionlistentry{}\label{fig:pa-li-nn-3d}
    \end{subfigure}\vspace{-20pt}\\

    \begin{subfigure}[t]{0.45\textwidth}
        \centering
        \includegraphics[width=\linewidth]{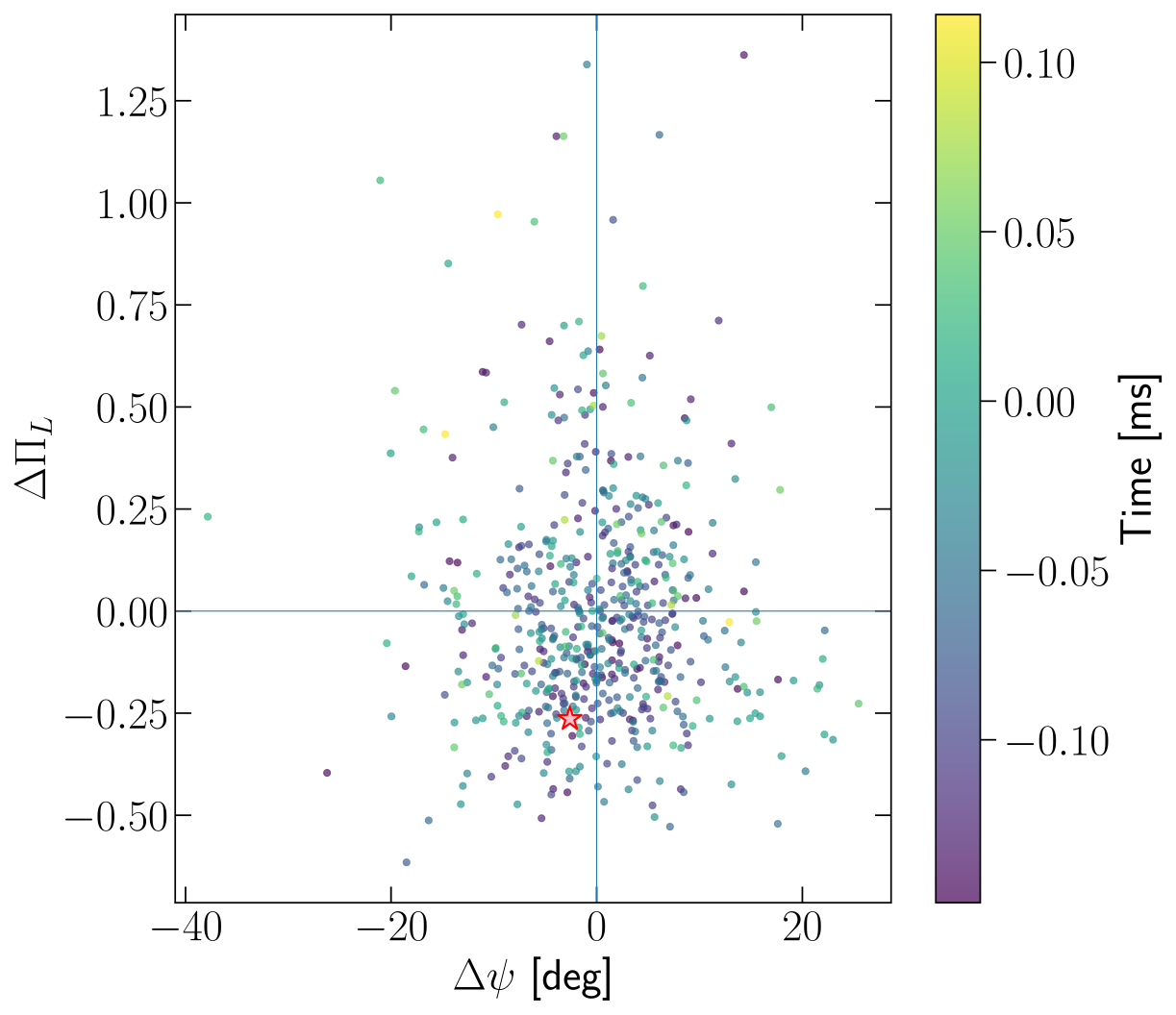}\par\medskip
        \captionlistentry{}\label{fig:pa-li}
    \end{subfigure}
    \begin{subfigure}[t]{0.45\textwidth}
        \centering
        \includegraphics[width=\linewidth, trim={0 0 0 1.5cm},clip]{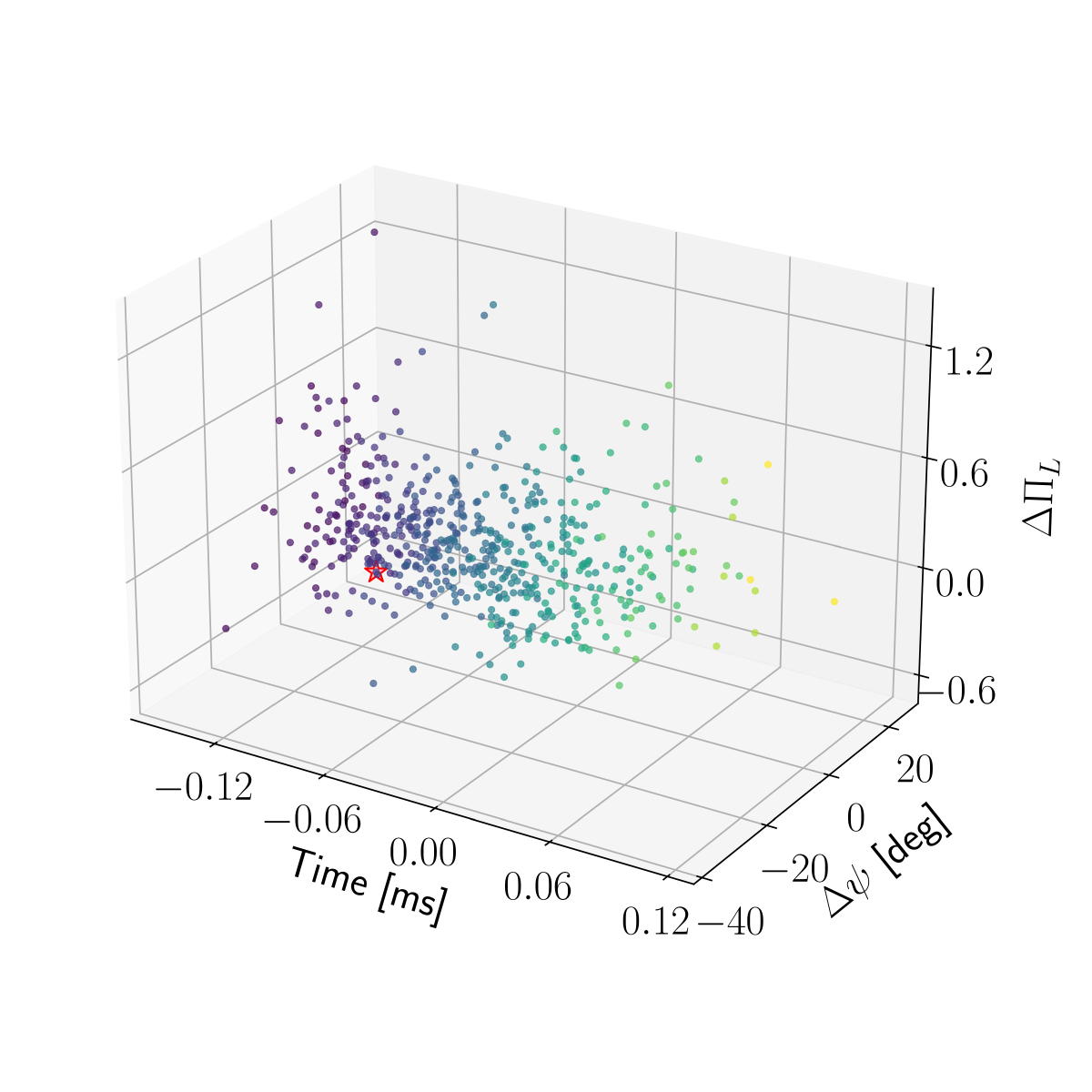}\par\medskip
        \captionlistentry{}\label{fig:pa-li-3d}
    \end{subfigure}

    \caption{Correlations between polarisation-angle and linear-polarisation fluctuations for mock FRB~20240318A, demonstrating noise-dominated behaviour.Left column: $\Delta\psi \equiv \psi-\bar{\psi}$ versus $\Delta\Pi_{L} \equiv \Pi_{L}-\bar{\Pi}_{L}$, with points coloured by time.Right column: corresponding 3D visualisation with time as the explicit axis. Rows (top to bottom): no noise, no scattering ($\tau_0=0$ ms); no noise, with scattering ($\tau_0=0.128$ ms); and with scattering plus realistic noise (S/N$\sim110$). We have increased the sampling rate $10\times$ compared to the main text to better inspect the tracks.}\label{fig:pa-li-full}
\end{figure*}

\begin{figure*}
    \addtocounter{figure}{-1}
    \renewcommand{\thefigure}{\arabic{figure} (continued)}
    \centering
    \begin{subfigure}[t]{0.45\textwidth}
        \centering
        \includegraphics[width=\linewidth]{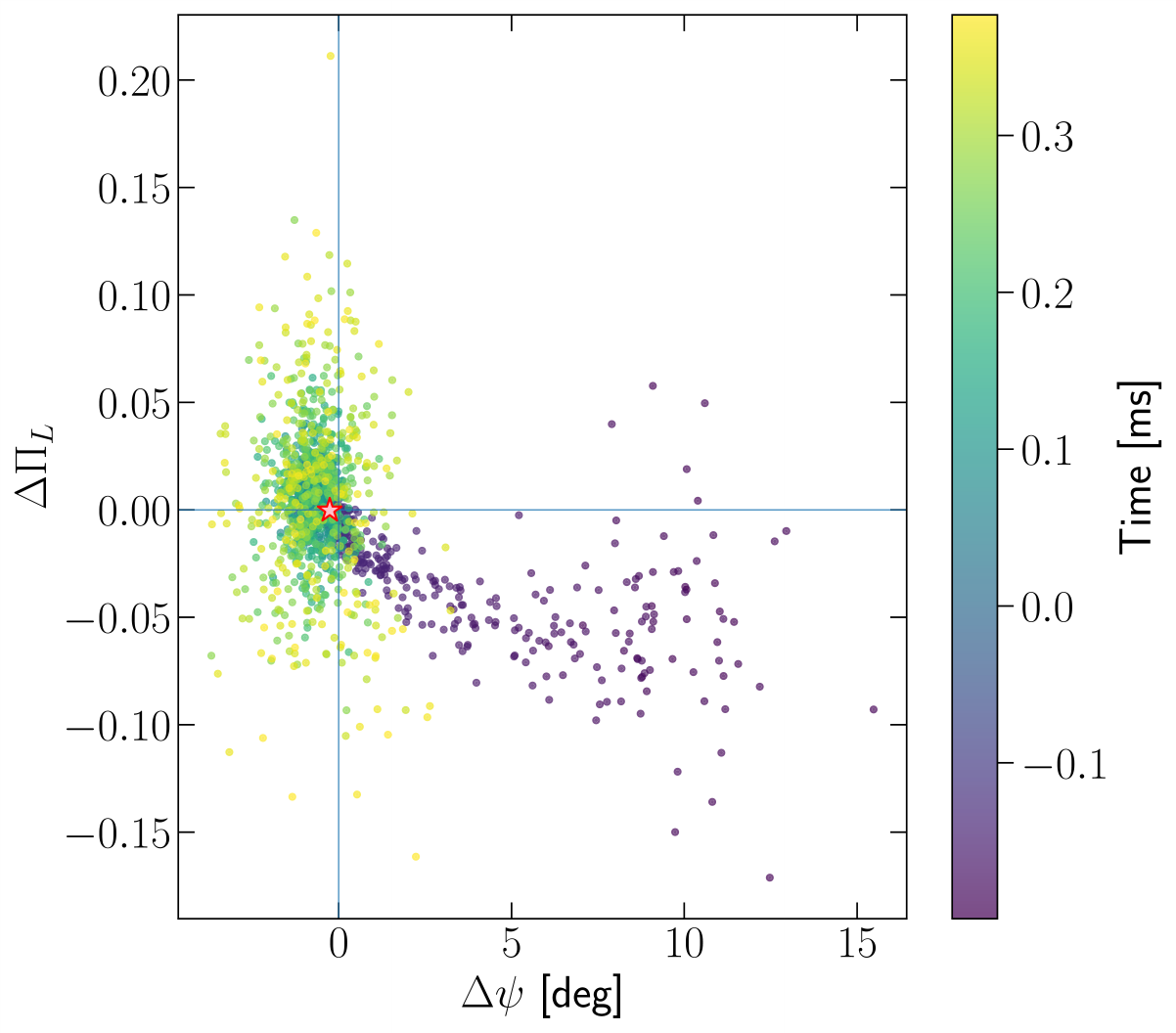}\par\medskip
        \captionlistentry{}\label{fig:pa-li_sn}
    \end{subfigure}
    \begin{subfigure}[t]{0.45\textwidth}
        \centering
        \includegraphics[width=\linewidth, trim={0 0 0 1.5cm},clip]{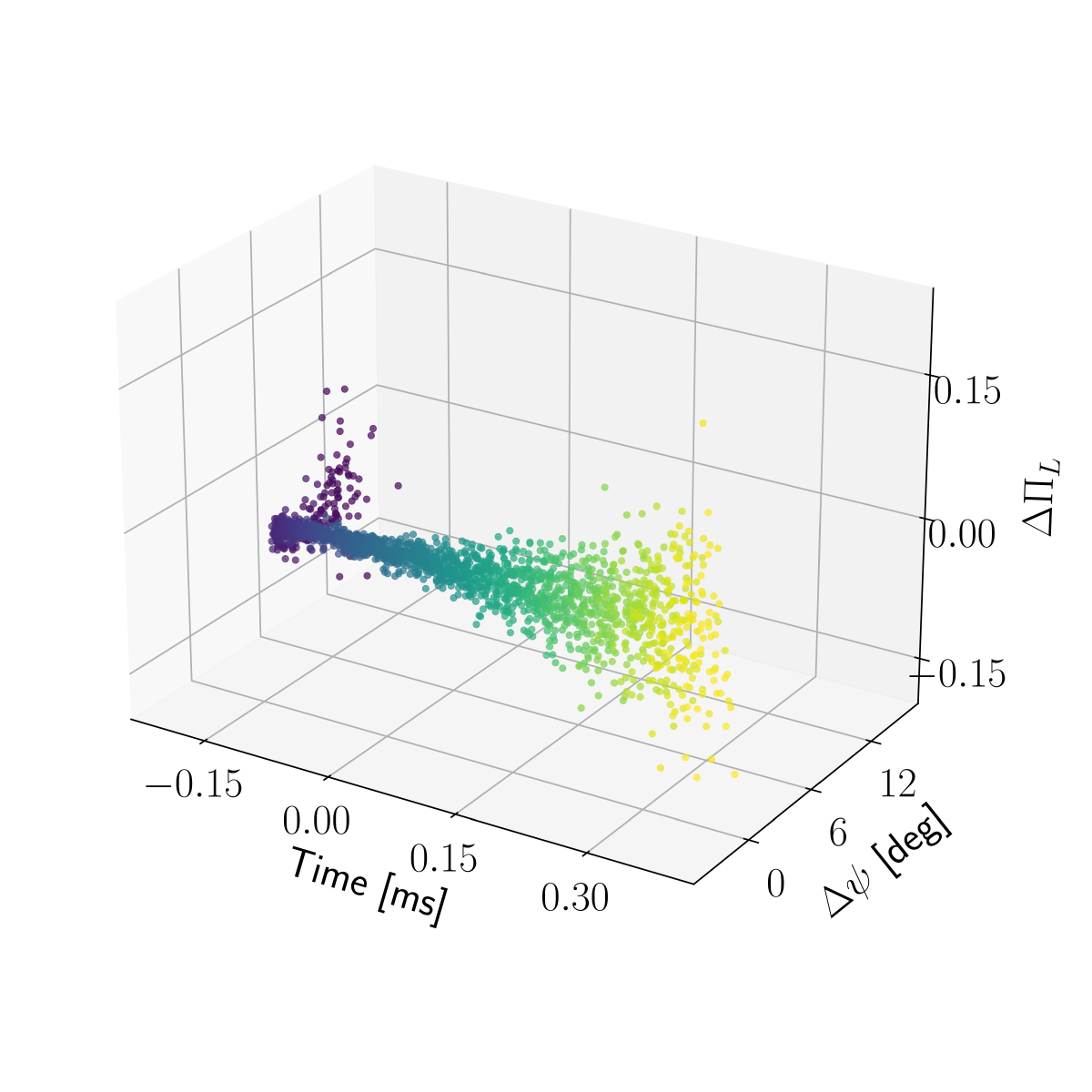}\par\medskip
        \captionlistentry{}\label{fig:pa-li_sn-3d}
    \end{subfigure}

    \begin{subfigure}[t]{0.45\textwidth}
        \centering
        \includegraphics[width=\linewidth]{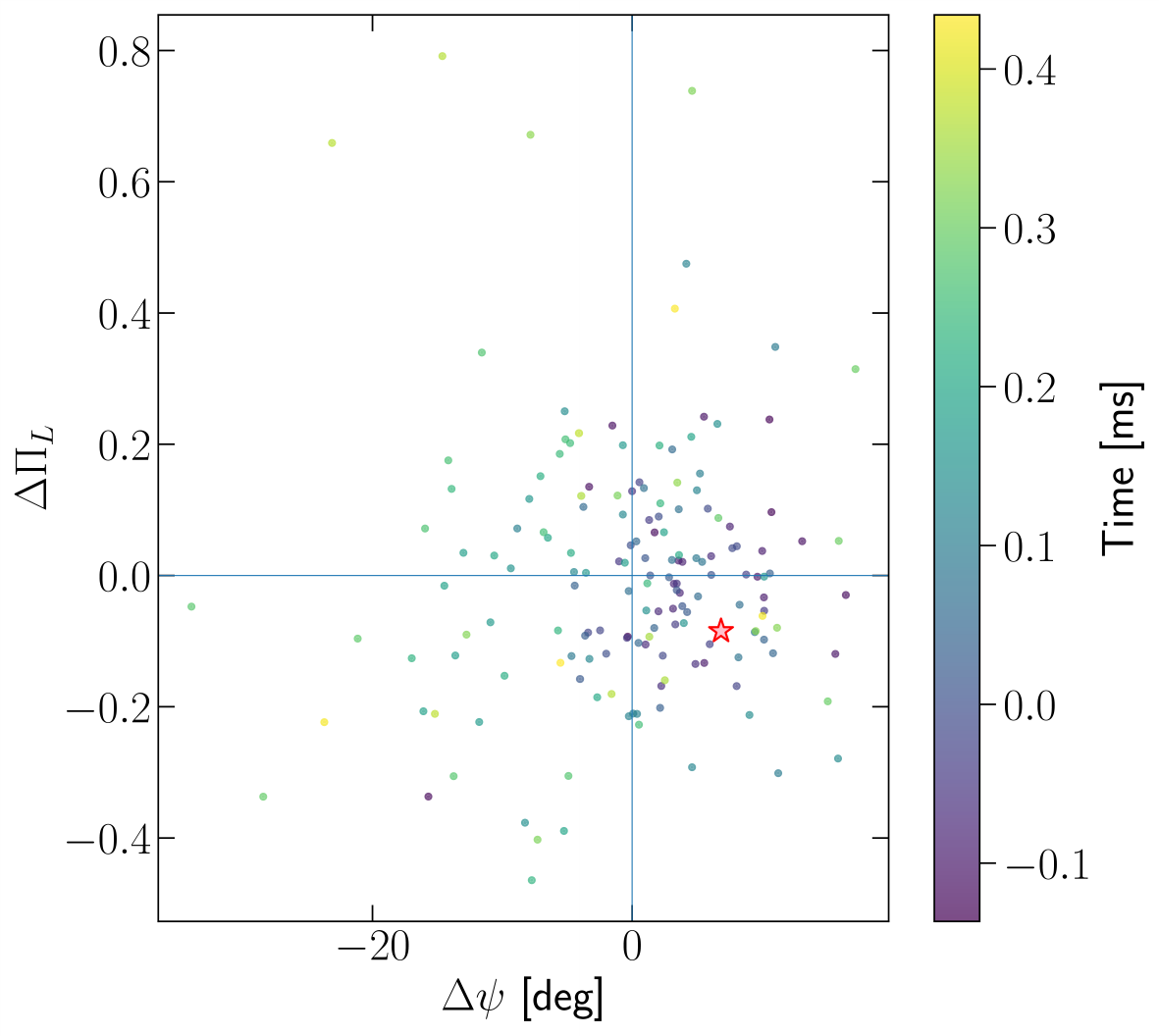}\par\medskip
        \captionlistentry{}\label{fig:pa-li-htr}
    \end{subfigure}
    \begin{subfigure}[t]{0.45\textwidth}
        \centering
        \includegraphics[width=\linewidth, trim={0 0 0 1.5cm},clip]{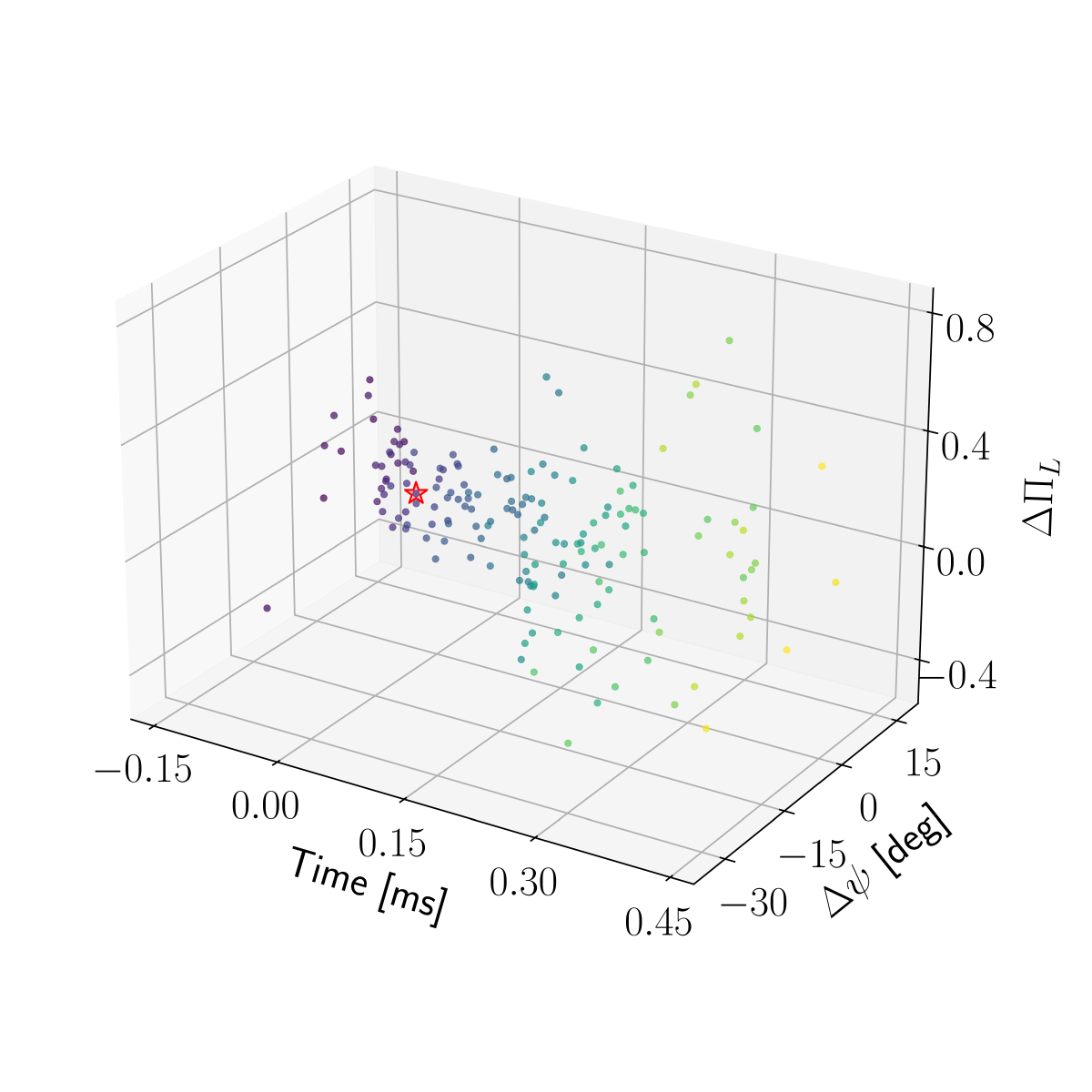}\par\medskip
        \captionlistentry{}\label{fig:pa-li-htr-3d}
    \end{subfigure}
    \caption{Top: scattering plus noise (S/N$\sim4900$), showing the correlation only becomes visible at unrealistically high sensitivity. Bottom: Real FRB~20240318A data from Figure~\ref{fig:HTR 240318A}.}
\end{figure*}
\renewcommand{\thefigure}{\arabic{figure}}

In the noiseless scattered case ($\tau_0=0.128$~ms), there is a clear anti-correlation, consistent with incoherence in $(Q,U)$ under strong overlap. The phase-space evolution in $(\Delta\psi,\Delta\Pi_L)$ is also consistent with this picture: as effective averaging increases across the burst, the point cloud contracts toward the mean state $(0,0)$, reflecting suppression of both large PA excursions and large linear-polarisation offsets. Early-time bins can deviate because intrinsic structure is less averaged (Section~\ref{subsec:PA variance}). By contrast, in the noiseless no-scattering baseline ($\tau_0=0$), the correlation is weaker, indicating that intrinsic PA spread alone is insufficient to guarantee strong $\Pi_L$ suppression. With realistic noise (S/N $\sim110$; $L$ masked where $I<2\sigma_I$), the relation disappears, implying that moderate intrinsic trends can be erased by measurement uncertainty and finite sensitivity. With noise added, the relation only begins to become visible at extremely high S/N.

Overall, the observed $\Delta\psi$--$\Delta\Pi_L$ behaviour is therefore controlled by a coupled parameter space --- intrinsic PA scatter, number of emitting elements, overlap fraction ($Nw_i/W_0$), scattering timescale, sampling interval, amplitude/width distributions, and S/N --- rather than any single parameter. This explains why flattened PA, PA variability, and low $\Pi_L$ need not map one-to-one across bursts, and why robust inference requires joint constraints on source microphysics and propagation/sensitivity effects (Section~\ref{subsec: depol degen}; cf. \citet{2025ApJ...979..160S,2025arXiv250517497S}).

\section{Impact of Spectral Modulation on PA Variance}\label{appendix:scint}

To assess whether fine spectral structure affects the inferred PA variance constraints, we repeated the mock-burst simulations shown in Figure~\ref{fig:LV} after applying representative scintillation modulation patterns to the spectra using the procedure described in Section~\ref{subsec: scint}. We found that the resulting $(\Pi_L,\mathbb{V}(\psi))$ relations were visually indistinguishable from those obtained without scintillation, with no systematic shifts apparent beyond the intrinsic simulation scatter.

This behaviour is expected because any diffractive scintillation pattern is effectively static over the duration of an FRB burst --- for Galactic scintillation at GHz frequencies, the characteristic scintillation timescale is many orders of magnitude longer than the millisecond burst durations considered here. While scintillation modulates the burst spectrum and local S/N, it cannot directly generate coherent time-dependent PA variations. We note that in principle, scintillation could indirectly affect PA results by introducing S/N-weighted frequency gaps that bias RM estimation; however, the RMs and scintillation strengths of the bursts in our sample are insufficient for this to be a concern.

While self-noise contributes to the uncertainty of spectral estimates, it does not produce correlated spectral modulation; instead, it manifests as uncorrelated variance in intensity estimates that averages down with the number of independent voltage samples within each integration. Stochastic wide-band impulse modulated self-noise \citep[SWIMS;][]{2011MNRAS.418.1258O} can produce temporally correlated noise variance when subpulse structure is resolved; however, since CRAFT data are Nyquist-sampled at 3~ns~\citep{2025PASA...42...36S}, yielding $> 10^3$ independent voltage samples per time bin at the time resolutions used here, any such cross-bin correlations in the Stokes parameters are significantly suppressed. Consequently, neither interstellar scintillation nor self-noise can generate the coherent time-dependent PA structure analysed in this work; self-noise may only contribute stochastic scatter to PA estimates.

\end{document}